\documentclass[12pt]{iopjournal}
\usepackage[T1]{fontenc}
\usepackage[utf8]{inputenc}
\usepackage{lmodern}
\usepackage[a4paper,margin=25mm,headheight=15pt]{geometry}
\usepackage{amsmath,amssymb,amsthm,mathtools}
\usepackage{booktabs,array}
\usepackage[numbers,sort&compress]{natbib}
\usepackage[font=small,labelfont=bf]{caption}
\usepackage{microtype}
\hypersetup{colorlinks=true,allcolors=black,
 pdftitle={Dual Affine Connections, Legendre Transforms, and Black Hole Thermodynamics},
 pdfauthor={Shoshauna Gauvin},pdfsubject={Black-hole thermodynamics and dual affine geometry}}
\fancyhfoffset{0pt}
\renewcommand{\articletype}[1]{}
\newcommand{\Cubic}{\mathsf C}
\newcommand{\Jvar}{\mathcal J}

\newcommand{\dd}{\mathrm d}
\newcommand{\LC}{\mathrm{LC}}
\newcommand{\R}{\mathbb R}
\newcommand{\E}{\mathbb E}
\DeclareMathOperator{\Var}{Var}
\DeclareMathOperator{\Cov}{Cov}
\DeclareMathOperator{\Tr}{Tr}
\DeclareMathOperator{\TV}{TV}
\DeclareMathOperator{\osc}{osc}

\newtheorem{theorem}{Theorem}[section]
\newtheorem{proposition}[theorem]{Proposition}

\newtheorem{corollary}[theorem]{Corollary}
\theoremstyle{definition}

\theoremstyle{remark}
\newtheorem{remark}[theorem]{Remark}
\numberwithin{equation}{section}

\begin{document}
\articletype{}
\title{Dual Affine Connections, Legendre Transforms, and Black Hole Thermodynamics}
\author{Shoshauna Gauvin}\par
\affil{University of Waterloo, Waterloo, Ontario, Canada}\par
\email{s3gauvin@uwaterloo.ca}\par
\vspace{3mm}

\begin{abstract}
We bound finite quantum information loss and black-hole entropy
increments using a positive scalar response and its variation.
On Hessian branches, the integrated cubic norm equals the total
variation of logarithmic response. We derive an explicit optimal
asymmetry bound under simultaneous response-range and affine
logarithmic-slope constraints, including nonmonotone responses;
full Legendre duality gives an invariant certificate.
For nested intervals in a thermal two-dimensional conformal field theory,
known infinite-line modular data express restriction loss as an exact Bregman
divergence in temperature squared. We prove complete monotonicity
and strict log-convexity of its response. At any fixed positive
temperature-squared step, the directed asymmetry uniquely identifies
the thermal offset within this family, independently of central
charge. This yields attained inverse intervals, a consistency test
across accessible fractions and certified Legendre-dual budgets.
A benchmark improves the range-only bound by $24.6\%$ and the
slope-only bound with the same budget by $2.73\%$. The semiclassical
planar Ba\~nados-Teitelboim-Zanelli interpretation concerns exterior thermal scales.
For uncertain thermodynamic responses, classical expectation
optimization gives sharp entropy constraints at fixed energy
increment. In an asymptotically flat Einstein-Maxwell spacetime with a finite cavity, we prove a uniform
response envelope for $|q|/R\le0.1$ on a specified matched large-horizon
branch. The exact family gives charge ordering, unique
charge-magnitude inference and an entropy test that rejects a
response-band-compatible pair with admissible energy. Sharpness
concerns the stated response classes; no microstate or dissipative
dynamics is inferred.
\end{abstract}

\section{Introduction}
\label{sec:introduction}

Restricting quantum access reduces the distinguishability of
equilibrium states. A related thermodynamic question asks which
finite entropy increments are compatible with an energy increment
and incomplete susceptibility data. Both comparisons can be expressed as
complementary Bregman remainders of a positive scalar response.
We derive a sharp bound from its range and affine logarithmic slope,
then apply it to thermal conformal field theory and a black hole
in a finite cavity.

The Hessian cubic measures the difference between dual affine
connections \cite{amari2000,shima2007}. In one dimension its integrated
norm is the Legendre-invariant total variation of logarithmic response.
Our principal geometric result is an explicit optimal asymmetry bound
under simultaneous range and affine logarithmic-slope constraints,
including nonmonotone responses. Combining both affine descriptions
gives an invariant certificate. The separate limits have antecedents
in approximate Bregman symmetry \cite{kindermann2019}, the
piecewise-quadratic square-root coefficient \cite{nilsson2026},
and bounded-log-derivative expectation optimization
\cite{walley1997,benavoli2015}.

For nested intervals in an unboosted thermal CFT$_2$ on the infinite
line, known modular expectations are affine in temperature squared
\cite{cardytonni2016,deboer2021}. They give an exact intrinsic Bregman
contrast without requiring an affine density-matrix curve.
We prove strict log-convexity of the loss response: at a known positive
temperature-squared increment, the directed asymmetry determines the
thermal offset within this family, independently of central-charge amplitude. The monotone
inverse gives attained uncertainty intervals and a consistency test
across accessible fractions. Certified bounds use both affine
descriptions. Under the semiclassical Einstein-AdS$_3$ dictionary,
the inferred scales have the standard planar-BTZ exterior
interpretation \cite{blanco2013,caggioli2024}. Thermal relative entropy
and the quadratic canonical-energy relation \cite{lashkari2016} are
established inputs; higher modular responses also have antecedents
\cite{sarosi2018}. Our additional results concern finite loss inference
and response bounds.

For uncertain thermodynamic responses, a reference shape gives sharp
bounds through classical expectile optimization
\cite{bellini2014,delbaen2013}. Fixing the energy increment adds a
moment constraint; the bathtub principle \cite[theorem 1.14]{liebloss2001}
then gives sharp entropy bounds and an inverse consistency test.
We apply these established optimization methods to Schwarzschild
and fixed-charge Reissner-Nordstr\"om cavities at common wall radius
and proper temperature. On a specified uniquely matched large-horizon
branch, we prove a one-sided response envelope within $2.983\%$ for
$|q|/R\le0.1$. Within the exact Einstein-Maxwell family, strict
ordering in charge squared instead lets energy identify charge
magnitude and predict entropy. An admissible-energy pair that fails
this entropy test exhibits information discarded by the envelope.
The equation of state and heat capacity are known
\cite{lundgren2008,wang2020,fernandes2025}; the new statements are
the uniform comparison and finite-increment inference. Envelope
optima need not occur in the smaller gravitational family.

Full Legendre duality relates descriptions of one model; it neither
identifies different ensembles nor supplies relative entropy between
fixed-charge sectors \cite{ghosh2019}. The cavity uses declared
classical saddle potentials, whereas the CFT provides an exact
quantum contrast. The entropy-geometry framework \cite{gauvin2026}
motivates the separation of statistical geometry, channels and
physical calibration.
All identities used here are proved below or attributed to
established literature.

Section~\ref{bnd:sec} derives the geometric bound and
section~\ref{qnt:sec} applies it to quantum interval loss.
Section~\ref{phys:sec} fixes the thermodynamic conventions for
response inference; section~\ref{cav:sec} gives the cavity application.
The appendices complete the optimization and gravitational proofs.

\section{Dual affine geometry}
\label{geo:sec}

Let $U\subset\R^n$ be open and convex, and let $\psi\in C^3(U)$ have
positive-definite Hessian in affine coordinates $\theta$:
\begin{equation}
 g_{ij}=\partial_i\partial_j\psi>0,\qquad
 \Cubic_{ijk}=\partial_i\partial_j\partial_k\psi.
 \label{geo:hessian}
\end{equation}
If $\nabla$ is the flat connection of $\theta$, its metric dual satisfies
\begin{equation}
 Xg(Y,Z)=g(\nabla_XY,Z)+g(Y,\nabla_X^*Z).
 \label{geo:duality}
\end{equation}
The standard Hessian identities \cite{amari2000,shima2007,nielsen2020} give
\begin{equation}
 \Gamma^k{}_{ij}=0,\qquad
 (\Gamma^*)^k{}_{ij}=g^{k\ell}\Cubic_{ij\ell},\qquad
 (\Gamma^{\LC})^k{}_{ij}=\tfrac12g^{k\ell}\Cubic_{ij\ell}.
 \label{geo:connections}
\end{equation}
Thus the Levi-Civita connection is their mean, and the affine
connections coincide on a connected domain exactly when $\Cubic=0$,
equivalently when $\psi$ is quadratic up to affine terms.
The cubic $\Cubic=\nabla g$ measures affine nonmetricity, not curvature
or energy.

The gradient map is a diffeomorphism onto its image. Full Legendre duality uses
\begin{equation}
 \eta_i=\partial_i\psi,\qquad
 \psi^*(\eta)=\theta^i\eta_i-\psi(\theta),\qquad
 \partial_{\eta_i}\partial_{\eta_j}\psi^*=g^{ij},
 \label{geo:legendre}
\end{equation}
and exchanges the affine roles while preserving the metric. In dual
coordinates,
\begin{equation}
 (\Cubic^*)^{ijk}=-g^{ia}g^{jb}g^{kc}\Cubic_{abc},\qquad
 \Cubic^*=\nabla^*g.
 \label{geo:dual-cubic}
\end{equation}
The sign reflects the exchanged connections; a coordinate change alone
does not change a geodesic's connection. Partial thermodynamic Legendre
transforms can change the ensemble, so their raw Hessians need not
reproduce the same positive metric \cite{ghosh2019}.

For a canonical Gibbs family with fixed Hamiltonian $H$,
$\psi=\log\Tr e^{-\beta H}$ and finite derivatives and moments,
\begin{equation}
 g=\psi''=\Var_\beta(H)=T^2C_{\rm heat},\qquad
 \Cubic=\psi'''=-\kappa_3(H),\qquad T=\beta^{-1}.
 \label{geo:canonical}
\end{equation}
For a gravitational saddle, $g$ and $\Cubic$ are derivatives of the
declared branch potential; their microscopic cumulant interpretation
requires control of that approximation.

\subsection{Quantum contrast and channel loss}
\label{geo:quantum-loss}

For faithful finite-dimensional states, set
$\Phi(\rho)=\Tr(\rho\log\rho)$ and
$D(\rho\Vert\sigma)=\Tr[\rho(\log\rho-\log\sigma)]$.
The trace-one constraint gives $D=B_\Phi$ in mixture-affine
coordinates, with the Bogoliubov-Kubo-Mori (BKM) Hessian metric
\cite{petz1994}. For a completely positive, trace-preserving $\Lambda$,
restrict outputs to the fixed support $\operatorname{supp}\Lambda(I)$.
Linearity gives
\begin{equation}
 \begin{gathered}
 \Delta_\Lambda(\rho,\sigma)
 :=D(\rho\Vert\sigma)-D(\Lambda\rho\Vert\Lambda\sigma)
 =B_{\Phi_\Lambda}(\rho,\sigma),\\
 \Phi_\Lambda(\rho)=\Phi(\rho)-\Phi(\Lambda\rho).
 \end{gathered}
 \label{geo:channel-bregman}
\end{equation}
The second term's derivative is composed with $\Lambda$, making its
Bregman remainder the output relative entropy. Data processing
\cite{lindblad1975} makes $\Phi_\Lambda$ convex, with Hessian and cubic
\begin{equation}
 g_{\rm loss}=g_{\rm BKM}-\Lambda^*g_{\rm BKM}\ge0,\qquad
 \Cubic_{\rm loss}=\nabla^{\rm m}g_{\rm loss},
 \label{geo:channel-jets}
\end{equation}
where the output metric is evaluated at $\Lambda\rho$ and
$\nabla^{\rm m}$ is the mixture connection. The cubic has no positivity
sign. On an affine faithful curve $\rho(u)=\rho_0+uV$, the loss is
$B_\phi$, $\phi=\Phi_\Lambda\circ\rho$; the scalar bounds below require
$\phi''>0$ throughout the comparison interval, excluding degenerate loss.
See also \cite[corollary 12]{gauvin2026}.

For faithful finite-dimensional states, $D(\rho\Vert\sigma)$ is the
optimal asymptotic type-II error exponent at a fixed type-I tolerance
in $(0,1)$ under collective tests on independent copies \cite{ogawa2000}.
Thus $\Delta_\Lambda$ is the exponent lost under the channel. The
copy-number limit uses a fixed state space; it supplies neither a
finite-copy bound nor an exchange with a continuum regulator limit.

Nonlinear state families need not inherit this affine Hessian
description. Section~\ref{qnt:sec} derives it from exact thermal CFT
modular data. A sharp continuum local algebra is not assumed to admit
a finite-dimensional state space or BKM construction without a faithful
differentiable realization.

\section{Integrated cubic response and directed divergence}
\label{bnd:sec}

Let $a<b$, $\psi\in C^3$ on a neighbourhood of $[a,b]$, and
$g=\psi''>0$. In one affine coordinate the cubic norm and metric measure
are
\begin{equation}
 \|\Cubic\|_g=\frac{|g'|}{g^{3/2}},\qquad
 \dd\mu_g=\sqrt g\,\dd u.
 \label{bnd:measure}
\end{equation}
For any positive constant reference $g_{\rm ref}$ of the same units,
their dimensionless integral is the total variation of logarithmic response:
\begin{equation}
 \Jvar_{[a,b]}:=\int_a^b\|\Cubic\|_g\,\dd\mu_g
 =\int_a^b\frac{|g'|}{g}\,\dd u
 =\TV_{[a,b]}\log\frac{g}{g_{\rm ref}}.
 \label{bnd:variation}
\end{equation}
Under full Legendre duality, $\dd\eta=g\,\dd u$ and $g^*=1/g$, hence
\begin{equation}
 \frac{|\partial_\eta g^*|}{g^*}\,\dd\eta
 =\frac{|g'|}{g}\,\dd u.
 \label{bnd:invariance}
\end{equation}
Thus $\Jvar$ is invariant on corresponding dual intervals. It equals
$|\log[g(b)/g(a)]|$ for monotone $g$ and accumulates all variations
otherwise; it is not metric length or an energy.

Define the Bregman divergence and its Legendre reversal
\cite{bregman1967,kindermann2019} by
\begin{align}
 B_\psi(x,y)&=\psi(x)-\psi(y)-\psi'(y)(x-y),
 \label{bnd:bregman}\\
 B_{\psi^*}(\eta_y,\eta_x)&=B_\psi(x,y),\qquad\eta_x=\psi'(x).
 \label{bnd:reversal}
\end{align}
For a fixed-Hamiltonian Gibbs family,
$B_\psi(b,a)=D(\rho_a\Vert\rho_b)$. Put $\Delta=b-a$ and
$h(t)=g(a+t\Delta)$. Taylor's integral remainder gives
\begin{equation}
 B_\psi(b,a)=\Delta^2\!\int_0^1(1-t)h(t)\,\dd t,\qquad
 B_\psi(a,b)=\Delta^2\!\int_0^1t h(t)\,\dd t.
 \label{bnd:remainders}
\end{equation}
The scale-invariant normalized asymmetry is therefore
\begin{equation}
 \mathcal A[h]
 =\frac{|B_\psi(b,a)-B_\psi(a,b)|}{B_\psi(b,a)+B_\psi(a,b)}
 =\frac{|\int_0^1(1-2t)h(t)\,\dd t|}{\int_0^1h(t)\,\dd t}.
 \label{bnd:slope-asymmetry}
\end{equation}
This compares the weighted mean of $t$ with the interval midpoint: a nonconstant
symmetric response has zero endpoint asymmetry despite positive $\Jvar$.

\subsection{The optimal joint comparison}
\label{bnd:joint-sec}

Bounded-log-derivative expectation models have piecewise-exponential
extrema \cite{walley1997,benavoli2015}. Adding a range constraint gives
the following comparison. Bounds on a density and its ordinary
derivative instead give plateau-linear extremizers \cite{kozine2009}.

\begin{theorem}[Joint response comparison]
\label{bnd:joint-theorem}
Let $h>0$ on $[0,1]$, let $\log h$ be Lipschitz, and suppose
\begin{equation}
 \osc\log h=\log\frac{\max h}{\min h}\le\omega,\qquad
 |(\log h)'|\le\kappa\ \hbox{almost everywhere},\qquad
 0\le\omega\le\kappa.
 \label{bnd:joint-hypotheses}
\end{equation}
Then $\mathcal A[h]\le\mathcal F(\omega,\kappa)$, where
$\mathcal F(0,\kappa)=0$ and, for $0<\omega\le\kappa$,
\begin{equation}
 \mathcal F(\omega,\kappa)
 =\coth\!\left(\frac\omega2\right)
 -\frac{\sqrt{1+\{4\sinh^2(\omega/2)-\omega^2\}/\kappa^2}}
 {\sinh(\omega/2)}.
 \label{bnd:joint-formula}
\end{equation}
This maximum is also the supremum over smooth positive responses
with the same budgets. For $\omega>0$, up to scaling
and reflection, the unique maximizer is
\begin{equation}
 h_{s_*}(t)=
 \begin{cases}
 e^\omega,&0\le t\le s_*,\\
 e^{\omega-\kappa(t-s_*)},&s_*\le t\le s_*+\omega/\kappa,\\
 1,&s_*+\omega/\kappa\le t\le1.
 \end{cases}
 \label{bnd:joint-extremizer}
\end{equation}
At the placement $s_*$ in \eqref{bnd:joint-root}, the ramp and full
response have the same weighted mean of $t$. If $0<\omega<\kappa$,
both plateaus have positive length and no $C^1$ response attains the maximum. If $\omega=\kappa>0$,
$s_*=0$ and the maximizers are the two exponentials; if $\omega=0$,
every admissible response is constant.
\end{theorem}

Appendix~\ref{app:joint-proof} gives the placement and the global
covariance certificate, without assuming a monotone competitor.
Nested admissible classes make $\mathcal F$ nondecreasing in each
budget. Its boundary values are
\begin{equation}
 \begin{gathered}
 \mathcal F(\omega,\omega)=\mathcal L(\omega),\qquad
 \lim_{\kappa\to\infty}\mathcal F(\omega,\kappa)=\tanh(\omega/4),\\
 \mathcal L(k)=\coth(k/2)-2/k\quad(k>0),\qquad\mathcal L(0)=0.
 \end{gathered}
 \label{bnd:joint-limits}
\end{equation}
For $0<\omega<\kappa<\infty$, the joint bound is strictly smaller
than both $\mathcal L(\kappa)$ and $\tanh(\omega/4)$.

\begin{corollary}[Separate range and slope comparisons]
\label{bnd:main}
\label{bnd:slope-theorem}
Under the standing interval hypotheses, let $m=\min g$,
$M=\max g$ and $\Jvar=\Jvar_{[a,b]}$. Then
\begin{equation}
 \sqrt{m/M}\le\frac{B_\psi(b,a)}{B_\psi(a,b)}\le\sqrt{M/m},
 \qquad \mathcal A[h]\le\tanh(\Jvar/4).
 \label{bnd:asymmetry-bound}
\end{equation}
The nonconstant range and variation constants are sharp suprema, not
smooth maxima. If $\log h$ is Lipschitz with slope budget $\kappa$,
then $\mathcal A[h]\le\mathcal L(\kappa)$; for $\kappa>0$, equality
holds exactly for $h=ce^{\pm\kappa t}$, $c>0$.
\end{corollary}

Every $C^1$ response has a finite slope budget. Letting the budget grow
in the joint theorem gives the range bound via
$\mathcal A=\tanh(|\log[B_\psi(b,a)/B_\psi(a,b)]|/2)$; use
$\log(M/m)\le\Jvar$ for the variation bound. Taking
$\omega=\kappa$ gives the slope comparison and its equality cases.
The joint maximizing sequences prove range and variation sharpness.
For continuous positive responses, extend by constant endpoint values
and mollify with a nonnegative smooth unit-mass kernel. The approximants
stay between $m$ and $M$ and converge uniformly, so both positive integrals
in \eqref{bnd:remainders} converge. This extends the range bound without
a derivative bound.
These limits recover the piecewise-quadratic square-root coefficient
\cite[remark 3.1]{nilsson2026} and exponential expectation comparison
\cite[section 7.4]{benavoli2015}.

\subsection{The Legendre-invariant certificate}
\label{bnd:joint-dual-sec}
\label{bnd:slope-dual-sec}

The smallest primal and dual affine slope budgets of $g$ are
\begin{equation}
 \kappa_{\rm p}=(b-a)\max\frac{|g'|}{g},\qquad
 \kappa_{\rm d}=\left(\int_a^b g\,\dd u\right)
                  \max\frac{|g'|}{g^2},
 \label{bnd:slope-dual}
\end{equation}
with maxima over $[a,b]$. The second follows from
$\partial_\eta\log g^*=-g'/g^2$. Conjugation preserves approximate symmetry \cite[lemma 2]{kindermann2019};
full Legendre duality exchanges these budgets and preserves $\omega=\log(M/m)$, $\Jvar$ and
$\mathcal A$, so
\begin{equation}
 0\le\omega\le\Jvar\le K:=\min\{\kappa_{\rm p},\kappa_{\rm d}\},
 \label{bnd:joint-hierarchy}
\end{equation}
and applying the theorem in both affine descriptions gives
\begin{equation}
 \mathcal A[h]\le\mathcal F(\omega,K)
 \le\mathcal F(\Jvar,K)\le\tanh(\Jvar/4).
 \label{bnd:joint-dual}
\end{equation}
Nested range constraints give the middle inequality; monotone responses
have $\omega=\Jvar$. The smaller affine budget retains the smooth sharp
supremum at fixed $0<\omega\le K$ (appendix~\ref{app:joint-proof}),
with the constant case $\omega=K=0$ trivial, without prescribing
an independent value of $\Jvar$.
Sharpness here concerns positive response geometry, without asserting
realizability by a fixed microscopic spectrum or gravitational theory.

\section{Quantum information loss for nested thermal intervals}
\label{qnt:sec}

The response bound also controls distinguishability lost as an observer's
accessible region shrinks. Consider a unitary nonchiral CFT$_2$ with equal
left and right central charges $c>0$ on the infinite spatial line, in
homogeneous thermal states without momentum or chemical potential.
Fix concentric intervals $I_l\subset I_L$, $0<l<L$, with a common,
temperature-independent endpoint UV prescription. Set $k_B=\hbar=1$,
$u=(\pi LT)^2$ and $r=l/L\in(0,1)$, using $u$ as the dimensionless affine
parameter. Finite spatial circles and disconnected intervals require
different modular data.

\subsection{Exact loss potential and its response}
\label{qnt:potential-sec}

For an interval of length $\xi L$, $\xi\in\{r,1\}$, remove its
state-independent entropy and define
\begin{equation}
 F_\xi(u)=-\frac c3\log\frac{\sinh(\xi\sqrt u)}{\xi\sqrt u},
 \qquad F_\xi(0)=0.
 \label{qnt:potential}
\end{equation}
The known thermal interval entropy and local modular Hamiltonian
\cite{cardytonni2016,deboer2021} give, at fixed reference $a>0$,
\begin{equation}
 \Delta S_\xi=-F_\xi(b)+F_\xi(a),\qquad
 \Delta\langle K_{\xi,a}\rangle
 =\frac c6\left(\frac ba-1\right)
       (\xi\sqrt a\coth(\xi\sqrt a)-1)
 =-F_\xi'(a)(b-a).
 \label{qnt:modular-affine}
\end{equation}
The second identity integrates the fixed modular weight against the
homogeneous energy density, proportional to $T^2$
\cite[section 7.2]{deboer2021}; its $a=0$ limit is $c\xi^2b/18$.
Endpoint constants and the UV divergence cancel in
$D=\Delta\langle K_a\rangle-\Delta S$, giving the restricted-state relative
entropy \cite[equation (7.27)]{deboer2021}, including its continuous vacuum limit,
\begin{equation}
 D_\xi(b\Vert a)=B_{F_\xi}(b,a),\qquad a,b\ge0.
 \label{qnt:exact-bregman}
\end{equation}
This intrinsic Bregman identity in $u$ requires neither affine density
matrices nor a Gibbs family with fixed modular Hamiltonian. In the
continuum, these finite differences evaluate the local-algebra relative
entropy $D_\xi$.

Define the directed losses and their potential by
\begin{equation}
 \begin{gathered}
 \Delta_+(a,b)=D_1(b\Vert a)-D_r(b\Vert a)=B_H(b,a),\\
 \Delta_-(a,b)=D_1(a\Vert b)-D_r(a\Vert b)=B_H(a,b),\qquad
 H=F_1-F_r,\quad h=H''.
 \end{gathered}
 \label{qnt:loss}
\end{equation}
With fixed endpoints $(a,b)$ suppressed, these measure thermal
distinguishability lost under restriction: a channel with a suitable
regulator, or restriction to a smaller local algebra in the continuum.
Their scalar response is
\begin{equation}
 h(u)=\frac c3\sum_{n=1}^{\infty}
 \left[(u+\pi^2n^2)^{-2}-(u+\pi^2n^2/r^2)^{-2}\right]>0 .
 \label{qnt:response}
\end{equation}
Twice differentiating Euler's product for $\sinh$ gives
$F_\xi''=(c/3)\sum_{n\ge1}(u+\pi^2n^2/\xi^2)^{-2}$,
with locally uniform convergence of all required derivatives on
$u\ge0$ and analytic extension around $u=0$. Thus $h$ is the difference
of the intrinsic relative-entropy metrics, with affine cubic response
$h'$. In a differentiable faithful density-matrix realization the metric
is the BKM pullback, rather than the symmetric-logarithmic-derivative
Fisher metric.

\begin{proposition}[Strict calibration and finite loss bound]
\label{qnt:main}
The response $h$ is completely monotone and strictly log-convex.
Its normalized cubic
\begin{equation}
 \mathcal R(u)=-\frac{h'(u)}{h(u)}
 \label{qnt:R}
\end{equation}
is independent of $c$ and is a continuous, strictly decreasing
bijection from $[0,\infty)$ onto $(0,\mathcal R(0)]$, where
\begin{equation}
 \mathcal R(0)=\frac4{21}\frac{1-r^6}{1-r^4},\qquad
 \mathcal R(u)\sim\frac3{2u}\quad(u\to\infty).
 \label{qnt:R-limits}
\end{equation}
For $0\le a<b<\infty$, put
\begin{equation}
 \omega=\log\frac{h(a)}{h(b)},\qquad
 \kappa=(b-a)\mathcal R(a).
 \label{qnt:budgets}
\end{equation}
Then $0<\omega<\kappa$, and the joint comparison of
theorem~\ref{bnd:joint-theorem} gives
\begin{equation}
 0<\mathcal A_{1\to r}(a,b)
 :=\frac{\Delta_+-\Delta_-}{\Delta_++\Delta_-}
 \le\mathcal F(\omega,\kappa),
 \qquad
 1<\frac{\Delta_+}{\Delta_-}
 \le\frac{1+\mathcal F(\omega,\kappa)}
          {1-\mathcal F(\omega,\kappa)}.
 \label{qnt:bound}
\end{equation}
\end{proposition}
\begin{proof}
Using $z^{-2}=\int_0^\infty t e^{-zt}\,\dd t$ in
\eqref{qnt:response} gives
\begin{equation}
 \begin{split}
 h(u)&=\int_0^\infty e^{-ut}\,\dd\mu(t),\\
 \dd\mu(t)&=\frac c3t\sum_{n\ge1}
       (e^{-\pi^2n^2t}-e^{-\pi^2n^2t/r^2})\,\dd t .
 \end{split}
 \label{qnt:laplace}
\end{equation}
The measure has positive density for every $t>0$ and finite moments
of every nonnegative integer order, proving complete monotonicity.
For the probability measure $e^{-ut}\dd\mu(t)/h(u)$,
$\mathcal R(u)$ is the mean of $t$ and $\mathcal R'(u)=-\Var_u(t)<0$.
This also proves strict log-convexity. The zero-temperature values
$h(0)=c(1-r^4)/270$ and $h'(0)=-2c(1-r^6)/2835$
follow from $\zeta(4)$ and $\zeta(6)$.
At infinity,
$H(u)=-c(1-r)\sqrt u/3+(c/3)\log(1/r)$ up to exponentially
small terms, giving $h(u)\sim c(1-r)/(12u^{3/2})$ and
\eqref{qnt:R-limits}. The limits and strict monotonicity prove bijectivity.

In the complementary kernels \eqref{bnd:remainders} for $H$, strict
decrease of $h$ and pairing $t$ with $1-t$ give
$\Delta_+-\Delta_->0$; both losses are positive.
Moreover $\omega=\int_a^b\mathcal R(u)\,\dd u<(b-a)\mathcal R(a)$, and
$\kappa$ is the exact affine logarithmic-slope maximum of
$t\mapsto h(a+t(b-a))$. Theorem~\ref{bnd:joint-theorem} proves
\eqref{qnt:bound}.
\end{proof}

Full Legendre duality can further reduce the certificate. With
$\Gamma=H'(b)-H'(a)>0$, equations~\eqref{bnd:slope-dual}-\eqref{bnd:joint-dual}
give
\begin{equation}
 \kappa_{\rm p}=(b-a)\mathcal R(a),\qquad
 \kappa_{\rm d}=\Gamma\max_{[a,b]}\frac{\mathcal R}{h},\qquad
 \mathcal A_{1\to r}\le\mathcal F(\omega,K),\quad
 K=\min\{\kappa_{\rm p},\kappa_{\rm d}\}.
 \label{qnt:dual}
\end{equation}
The dual maximum admits a certified enclosure: for any integer $N\ge1$,
set $u_j=a+j(b-a)/N$. Since both $h$ and $\mathcal R$ decrease,
\begin{equation}
 \begin{gathered}
 \kappa_{\rm d}\le
 \kappa_{{\rm d},N}:=\Gamma\max_{0\le j<N}
       \frac{\mathcal R(u_j)}{h(u_{j+1})}
 \le e^{\kappa_{\rm p}/N}\kappa_{\rm d},\\
 K_N=\min\{\kappa_{\rm p},\kappa_{{\rm d},N}\},\qquad
 \mathcal A_{1\to r}\le\mathcal F(\omega,K_N),\qquad
 K\le K_N\le e^{\kappa_{\rm p}/N}K .
 \end{gathered}
 \label{qnt:dual-mesh}
\end{equation}
Indeed, on a grid cell, $\mathcal R/h\le\mathcal R(u_j)/h(u_{j+1})$.
The latter is at most
$\max(\mathcal R/h)\exp(\int_{u_j}^{u_{j+1}}\mathcal R\,\dd u)
\le\max(\mathcal R/h)e^{\kappa_{\rm p}/N}$.
Thus the certified budgets converge to the Legendre-invariant $K$.

For $r=1/2$ and $[a,b]=[1/4,4]$, the $N=256$ certificate gives
\begin{equation}
 \begin{gathered}
 (\Delta_+,\Delta_-)/c
       \simeq(0.018790488381,\,0.015284588753),\\
 (\omega,\kappa_{\rm p},K_{256})
       \simeq(0.619228358,\,0.731057515,\,0.720521300),\\
 \mathcal A_{1\to r}\simeq0.102887504,\qquad
 \mathcal F(\omega,K_{256})\simeq0.115812602,\\
 \tanh(\omega/4)\simeq0.153582168,\qquad
 \mathcal L(K_{256})\simeq0.119060509 .
 \end{gathered}
 \label{qnt:benchmark}
\end{equation}
The joint certificate is $24.6\%$ below the range bound and $2.73\%$
below the slope bound using the same budget. It is also $0.919\%$
below the primal joint certificate $\mathcal F(\omega,\kappa_{\rm p})$.
Outward-rounded interval arithmetic certifies these rounded comparisons.
These exact losses benchmark certificates usable with only range and slope
information about the response.
Sharpness holds over the larger class in theorem~\ref{bnd:joint-theorem};
the CFT family need not attain it.

\subsection{What finite thermal contrasts calibrate in planar BTZ}
\label{qnt:gravity-sec}

At fixed interval ratio $r$ and calibrated step $\delta>0$, the normalized
loss
\[
 A_{r,\delta}(a)=\mathcal A_{1\to r}(a,a+\delta)
\]
identifies a thermal parameter at finite separation.

\begin{proposition}[Finite-step calibration and consistency]
\label{qnt:finite-calibration}
For each $r\in(0,1)$ and $\delta>0$, the map $A_{r,\delta}$ is independent
of $c$ and is a continuous, strictly decreasing bijection
\[
 [0,\infty)\longrightarrow(0,A_{r,\delta}(0)],
 \qquad A_{r,\delta}(a)=\frac{\delta}{4a}+O(a^{-2})
 \quad(a\to\infty).
\]
Writing $b=a+\delta$ and using \eqref{qnt:budgets} and \eqref{qnt:dual}, its
finite-step certificates satisfy
\begin{equation}
 \mathcal L\bigl(\delta\mathcal R(b)\bigr)
 < A_{r,\delta}(a)
 \le \mathcal F(\omega,K)
 < \mathcal L\bigl(\delta\mathcal R(a)\bigr).
 \label{qnt:two-sided}
\end{equation}
Consequently, a certified interval
$0<\alpha_-\le A_{r,\delta}(a)\le\alpha_+\le A_{r,\delta}(0)$
gives the exact attained model interval
\begin{equation}
 a\in\bigl[A_{r,\delta}^{-1}(\alpha_+),
           A_{r,\delta}^{-1}(\alpha_-)\bigr].
 \label{qnt:finite-inverse}
\end{equation}
\end{proposition}
\begin{proof}
Let $p_a(t)=h(a+\delta t)/\int_0^1h(a+\delta s)\,\dd s$ on $[0,1]$.
The complementary kernels give $A_{r,\delta}(a)=1-2\E_a[t]$.
Differentiation on this compact interval yields
\[
 A_{r,\delta}'(a)
 =-2\Cov_a
 \left(t,\frac{h'(a+\delta t)}{h(a+\delta t)}\right)<0.
\]
Strict log-convexity makes $q(t)=h'(a+\delta t)/h(a+\delta t)$ strictly
increasing. The covariance is half the expectation of
$(t-s)(q(t)-q(s))$, positive off the diagonal for independent draws.
Positivity follows from \eqref{qnt:bound}.
The asymptotic $h(u)\sim c(1-r)/(12u^{3/2})$, with exponentially small
relative correction, gives
$p_a(t)=1-\frac{3\delta}{2a}(t-\frac12)+O(a^{-2})$
uniformly in $t$, proving the limit and bijection.
For the lower bound, $h(a+\delta t)/e^{-\delta\mathcal R(b)t}$ is
strictly decreasing; the same covariance argument compares the mean under
$p_a$ with the normalized exponential's, whose asymmetry is
$\mathcal L(\delta\mathcal R(b))$.
The upper bounds follow from \eqref{qnt:dual}, $K\le\kappa_{\rm p}$
and the strict joint improvement at $\omega<\kappa_{\rm p}$.
Monotonicity and continuity then prove \eqref{qnt:finite-inverse},
including attainment of every parameter in that interval.
\end{proof}

At fixed $L$ and $\delta$, distinct accessible fractions $r_1,r_2$ test
consistency with the one-parameter thermal family. Write
$A_i=A_{r_i,\delta}$: an admissible first contrast $\alpha$ predicts
$A_2(A_1^{-1}(\alpha))$. Under the admissibility conditions of
proposition~\ref{qnt:finite-calibration},
$\alpha_-\le A_1(a)\le\alpha_+$ predicts exactly
\begin{equation}
 A_2(a)\in
 \left[A_2\!\left(A_1^{-1}(\alpha_-)\right),
       A_2\!\left(A_1^{-1}(\alpha_+)\right)\right].
 \label{qnt:cross-access}
\end{equation}
Every value, including both endpoints, is attained. A second certified
interval is compatible with the family and first interval exactly when
it intersects \eqref{qnt:cross-access}. Compatibility does not establish
uniqueness of the physical model.

For example, take $\delta=15/4$ and $(r_1,r_2)=(1/2,3/4)$.
The admissible first interval $A_1(a)\in[0.102,0.104]$ gives exact
attained intervals through \eqref{qnt:finite-inverse} and
\eqref{qnt:cross-access}, with outward-rounded numerical enclosures
\begin{equation}
 a\in[0.126560,\,0.350349],\qquad
 A_2(a)\in[0.117301,\,0.119555].
 \label{qnt:inverse-example}
\end{equation}
Directed interval arithmetic certifies both inverse endpoints and their
images; the displayed decimal endpoints are outer bounds, not attained
endpoints of the exact intervals. A second certified interval
$A_2(a)\in[0.120,0.1205]$ is individually admissible but incompatible
with this prediction.

Taylor expansion of $H$ recovers the normalized cubic:
\begin{equation}
 A_{r,\delta}(a)=\frac{\delta}{6}\mathcal R(a)+O(\delta^2),
 \label{qnt:local-calibration}
\end{equation}
These calibrations require independently known affine normalization of
$u$, increments $\delta$ and interval sizes. No acquisition protocol or
metrological advantage follows. Sensitivity decreases and the inverse
becomes poorly conditioned at large $a$, where
$A_{r,\delta}'(a)\sim-\delta/(4a^2)$.

For a CFT with a controlled classical Einstein-AdS$_3$ dual,
the thermal interval entropy is the leading order-$c$ RT entropy of
planar BTZ \cite{blanco2013}. With boundary metric $-\dd t^2+\dd x^2$
and AdS radius $\ell_{\rm AdS}$, choose the noncompact planar radial
coordinate with $g_{xx}=r_{\rm bulk}^2/\ell_{\rm AdS}^2$.
The corresponding horizon and
interval-$d$ turning radii are \cite{caggioli2024}
\begin{equation}
 r_h=\frac{2\ell_{\rm AdS}^2}{L}\sqrt u,\qquad
 r_*(d)=r_h\coth\!\left(\frac dL\sqrt u\right),
 \qquad r_*(d)\big|_{u=0}=\frac{2\ell_{\rm AdS}^2}{d}.
 \label{qnt:btz-scales}
\end{equation}
The finite-step calibration therefore fixes boundary-normalized thermal,
horizon and RT-depth scales within this family. Temperature changes
compare different backgrounds, with depths evaluated in each geometry.
The planar exterior is locally AdS$_3$, with scalar curvature
$-6/\ell_{\rm AdS}^2$ independent of $r_h$; the calibration does not
distinguish local curvature, compact quotient data, interiors or microstates.

At the vacuum endpoint, the established identification of the
relative-entropy Hessian with canonical energy
\cite[sections 3.4 and 4.3]{lashkari2016} gives, at leading classical order and
for the same normalized thermal perturbation,
\begin{equation}
 D_\xi(u\Vert0)=\tfrac12u^2\mathcal E_\xi+O(u^3),
 \qquad
 \mathcal E_1-\mathcal E_r=h(0)=\frac{c(1-r^4)}{270}.
 \label{qnt:canonical-energy}
\end{equation}
The wedge canonical-energy functionals have different modular generators
and surface terms; their difference is not automatically a local shell
energy. Finite comparisons use the exact boundary formula, not an exact
finite JLMS equality with order-one bulk matter relative entropy
\cite{jlms2016}. Holographic accuracy depends on the supplied dictionary;
lattice regulators need not preserve the exact conformal formulas.

From established thermal formulas and canonical-energy inputs, we have
proved positivity of the nested-loss response, exact finite-step inversion
and consistency tests, and certified bounds in both affine descriptions.
Higher modular responses occur in nonlinear holographic perturbation
theory \cite{sarosi2018,faulkner2017}. The cubic adds information beyond
the metric at one reference, not beyond the full metric field: it measures
a connection difference in this intrinsically flat scalar model.

\section{Thermodynamic interpretation and endpoint conventions}
\label{phys:sec}

We set the speed of light, $\hbar$ and $k_{\mathrm B}$ to unity and retain $G$;
$c$ in section~\ref{qnt:sec} denotes central charge. Gravitational responses
are derivatives of specified saddle potentials. Identifying them with exact
microscopic cumulants requires a supplied Gibbs model and controlled
saddle corrections.

Neither $\Jvar$ nor normalized asymmetry has an energy dimension.
A positive rescaling of the potential scales both divergences and the
calibrated energy, but preserves normalized comparisons.

\subsection{A common first-law endpoint identity}

Hold the other controls fixed and suppose
$\dd S=\beta\,\dd E$ on a regular interval with $\beta>0$ and
$g=-\dd E/\dd\beta>0$. For $\psi=S-\beta E$ and
$\beta_b<\beta_a$, write $\Delta E=E_b-E_a>0$ and
$\Delta S=S_b-S_a$. Then $\psi'=-E$, $\psi''=g$, and
\begin{subequations}\label{phys:canonical-endpoints}
\begin{align}
 B_\psi(\beta_b,\beta_a)&=\Delta S-\beta_b\Delta E,
 \label{phys:canonical-forward}\\
 B_\psi(\beta_a,\beta_b)&=\beta_a\Delta E-\Delta S.
 \label{phys:canonical-reverse}
\end{align}
\end{subequations}
Substitution in the Bregman formula gives these identities.
Both divergences are strictly positive because
$\Delta S=\int_{E_a}^{E_b}\beta(E)\,\dd E$ and
$\beta_b<\beta(E)<\beta_a$ in the interior. Their sum is
$(\beta_a-\beta_b)\Delta E$ and their signed difference is
$2\Delta S-(\beta_a+\beta_b)\Delta E$.
For a supplied canonical Gibbs family, the forward orientation is
$D(\rho_{\beta_a}\Vert\rho_{\beta_b})$. The cavity application below
uses the quasilocal energy at fixed wall radius and charge.

\subsection{Entropy increments from incomplete response information}
\label{phys:secant-sec}

Let $\Delta\beta=\beta_a-\beta_b>0$ and
$\bar\beta=(\beta_a+\beta_b)/2$. The signed physical comparison is
\begin{equation}
 \alpha:=\frac{B_\psi(\beta_b,\beta_a)-B_\psi(\beta_a,\beta_b)}
 {B_\psi(\beta_b,\beta_a)+B_\psi(\beta_a,\beta_b)}
 =\frac{2\Delta S-(\beta_a+\beta_b)\Delta E}
 {\Delta\beta\,\Delta E}.
 \label{phys:signed-comparison}
\end{equation}
Consequently a certificate $\alpha_-\le\alpha\le\alpha_+$ gives
\begin{equation}
 \left(\bar\beta+\frac{\alpha_-}2\Delta\beta\right)\Delta E
 \le\Delta S\le
 \left(\bar\beta+\frac{\alpha_+}2\Delta\beta\right)\Delta E.
 \label{phys:secant-certificate}
\end{equation}
An absolute bound $\mathcal A\le B$ supplies $\alpha_\pm=\pm B$.
For the ascending affine interval $[\beta_b,\beta_a]$ in
proposition~\ref{ctl:sharp-band}, $\alpha=-\sigma$, so its sharper
signed limits are $\alpha_-=-m_+$ and $\alpha_+=-m_-$.

Equation~\eqref{phys:secant-certificate} holds at supplied $\Delta E$,
although its sharpness may allow that increment to vary.
Both calibrated increments determine the ascending-interval response moments
\begin{equation}
 H=\frac{\Delta E}{\Delta\beta},\qquad
 M=\frac{(\beta_a+\beta_b)\Delta E-2\Delta S}{(\Delta\beta)^2},
 \qquad \alpha=-\frac MH.
 \label{ctl:calibrated-moments}
\end{equation}
Proposition~\ref{ctl:fixed-integral} gives sharp fixed-energy entropy bounds
by setting $\alpha_-=-M_+/H$ and $\alpha_+=-M_-/H$ in
\eqref{phys:secant-certificate}. Corollary~\ref{ctl:calibrated-inverse}
gives the exact infimum of response allowance required by both increments.
The normalized inverse \eqref{ctl:inverse-epsilon}, using only $M/H$,
remains necessary when additional data are fixed.

The remainders in \eqref{phys:canonical-endpoints} are finite Clausius
deficits; their endpoint values imply no thermalization protocol.
Using declared temperature and energy data with an explicit response
defect follows the input-and-calibration distinction of
\cite[definition 1 and section 3.5]{gauvin2026}; the identities here
follow directly from the first law.

\subsection{Certificates with uncertain susceptibility}
\label{ctl:transfer-sec}

Uniform susceptibility error transfers finite-separation certificates
without cubic-error control.

\begin{theorem}[Response-error transfer]
\label{ctl:transfer}
Let $\psi,\psi_0\in C^2$ on a neighbourhood of the same compact affine
interval, with $g=\psi''>0$ and $g_0=\psi_0''>0$. Suppose
\begin{equation}
 (1-\epsilon)g_0\le g\le(1+\epsilon)g_0,\qquad 0\le\epsilon<1.
 \label{ctl:hypothesis}
\end{equation}
For distinct endpoints, let $\mathcal A$ and $\mathcal A_0$ be their
normalized Bregman asymmetries. Then
\begin{equation}
 \mathcal A\le T_\epsilon(\mathcal A_0),\qquad
 T_\epsilon(B)=\frac{B+\epsilon}{1+\epsilon B}.
 \label{ctl:bound}
\end{equation}
Any certificate $\mathcal A_0\le B_0<1$ consequently gives
$\mathcal A\le T_\epsilon(B_0)<1$.
\end{theorem}

\begin{proof}
By the positive kernels in \eqref{bnd:remainders}, each orientation for
$g$ divided by its value for $g_0$ lies in $[1-\epsilon,1+\epsilon]$.
The ratios of orientations $R,R_0$ therefore satisfy
\begin{equation}
 |\log R|\le|\log R_0|+\log\frac{1+\epsilon}{1-\epsilon}.
 \label{ctl:ratio-proof}
\end{equation}
Now use $\mathcal A=\tanh(|\log R|/2)$, tanh addition, and
$T_\epsilon'(B)=(1-\epsilon^2)/(1+\epsilon B)^2>0$.
\end{proof}

For example, take $B_0=\mathcal F(\omega_0,K_0)$.
A target $\mathcal A\le A_{\max}$ with $B_0<A_{\max}<1$ is ensured by
\begin{equation}
 \epsilon\le\frac{A_{\max}-B_0}{1-A_{\max}B_0}.
 \label{ctl:accuracy}
\end{equation}
Optimality is not claimed when both kernels share one response error.
Comparison requires matched physical controls in the same declared affine
coordinate; independent Legendre transformations need not preserve this matching.

Small potential-value error is insufficient: $\alpha\cos(nu)$ has amplitude
$\alpha$ but second-derivative amplitude $\alpha n^2$.
Section~\ref{cav:rn-controlled} derives a uniform response envelope
for a specified classical matter family.

\paragraph{Direct range control and the value of slope information.}
For a band $l g_0\le g\le u g_0$ with $0<l\le u$, put
$\omega_0=\osc\log g_0$ and $\delta=\log(u/l)$.
Since $\osc\log g\le\omega_0+\delta$, the range bound directly gives
\begin{equation}
 \mathcal A[g]\le B_{\rm range}
 :=\tanh\frac{\omega_0+\delta}{4}.
 \label{ctl:direct-range}
\end{equation}
This improves the transferred range bound $\tanh(\omega_0/4+\delta/2)$
strictly when $l<u$, with equality otherwise.
If reference slope information supplies $F_0=\mathcal F(\omega_0,K_0)$,
recentring the band gives $e=(u-l)/(u+l)=\tanh(\delta/2)$.
The joint transfer is $T_e(F_0)=\tanh(\operatorname{arctanh}F_0+\delta/2)$;
therefore
\begin{equation}
 T_e(F_0)<B_{\rm range}
 \quad\Longleftrightarrow\quad
 \delta<\omega_0-4\operatorname{arctanh}F_0.
 \label{ctl:joint-advantage}
\end{equation}
This compares the two certificates without asserting joint optimality
under response error or assuming a target derivative bound.

\subsection{Exact uncertainty bounds with a known reference response}
\label{ctl:sharp-band-sec}

The full reference response gives a stronger certificate through the
expectile representation of robust expectations
\cite{bellini2014,delbaen2013}. The thermodynamic application uses the
common response error in both Bregman kernels to test their ordering.

For measurable responses, inequalities and uniqueness hold almost
everywhere; $\|\cdot\|_\infty$ is the essential-supremum norm, or the
uniform norm for continuous responses.

\begin{proposition}[Sharp comparison in a relative-response band]
\label{ctl:sharp-band}
Let $h_0>0$ be continuous on $[0,1]$ and $0\le\epsilon<1$. Put
\begin{equation}
 s(t)=1-2t,\quad H_0=\int_0^1h_0\,\dd t,\quad
 M_0=\int_0^1s h_0\,\dd t,\quad
 J_0(m)=\int_0^1|s-m|h_0\,\dd t.
 \label{ctl:band-moments}
\end{equation}
There are unique roots $m_-,m_+\in(-1,1)$ of
\begin{equation}
 M_0-m_\pm H_0\pm\epsilon J_0(m_\pm)=0.
 \label{ctl:band-roots}
\end{equation}
Every measurable response satisfying
$(1-\epsilon)h_0\le h\le(1+\epsilon)h_0$ obeys
\begin{equation}
 m_-\le \frac{\int_0^1s h\,\dd t}{\int_0^1h\,\dd t}\le m_+,
 \qquad
 \mathcal A[h]\le A_{\rm band}:=\max\{|m_-|,|m_+|\}.
 \label{ctl:band-sharp}
\end{equation}
Both signed endpoints are attained by measurable responses and are the infimum and
supremum over continuous responses (also smooth responses if $h_0$ is
smooth). Neither is attained by continuous responses for $\epsilon>0$. In particular,
$A_{\rm band}\le T_\epsilon(|M_0|/H_0)$.
\end{proposition}

Appendix~\ref{app:response-band} proves the result and evaluates its
endpoint moments.

The roots are the $(1\pm\epsilon)/2$ expectiles of $s(t)$ under
$h_0(t)\,\dd t/H_0$, computed by two scalar monotone searches.
For a constant reference,
\begin{equation}
 A_{\rm band}
 =\frac{\sqrt{1+\epsilon}-\sqrt{1-\epsilon}}
 {\sqrt{1+\epsilon}+\sqrt{1-\epsilon}},
 \label{ctl:constant-band}
\end{equation}
whereas \eqref{ctl:bound} gives $\epsilon$.

For the exact ordering tolerance, when $M_0\ne0$ define
\begin{equation}
 \epsilon_{\rm ord}:=\frac{|M_0|}{J_0(0)}\in(0,1).
 \label{ctl:ordering-tolerance}
\end{equation}
For $\epsilon<\epsilon_{\rm ord}$, all admissible responses retain
the reference ordering with a uniform positive signed margin.
For $\epsilon>\epsilon_{\rm ord}$, continuous responses can reverse it,
as can smooth responses if $h_0$ is smooth. At equality the measurable
extremizer has equal divergences; continuous responses retain strict
ordering without a uniform margin. This follows by minimizing the
reference-signed numerator, whose infimum is $|M_0|-\epsilon J_0(0)$,
and requires no bound on the unknown response derivative.

As in theorem~\ref{ctl:transfer}, both responses use the same physical
endpoints and declared affine coordinate. Sharpness is within this
response class; arbitrary members need not arise from a specified
microscopic spectrum or controlled gravitational corrections.

\begin{corollary}[Correction required by a normalized endpoint comparison]
\label{ctl:inverse-band}
In proposition~\ref{ctl:sharp-band}, the smallest uniform
relative-response allowance permitting signed mean $m\in(-1,1)$ is
\begin{equation}
 \epsilon_{\rm req}(m)
 =\frac{|mH_0-M_0|}{J_0(m)}<1.
 \label{ctl:inverse-epsilon}
\end{equation}
The measurable minimum is attained. For continuous responses (also
smooth responses if $h_0$ is smooth), it is the infimum, attained only
when $m=M_0/H_0$. Every larger allowance below one admits a response
of the corresponding regularity with mean exactly $m$.

Equivalently, the least oscillation of the logarithmic response ratio is
\begin{equation}
 \delta_{\rm req}(m)
 =2\operatorname{arctanh}\epsilon_{\rm req}(m),\qquad
 \delta=\osc\log(h/h_0),
 \label{ctl:inverse-shape}
\end{equation}
with the same attainment conventions. A positive constant calibration
of either response does not affect this second diagnostic.
\end{corollary}

The proof is in appendix~\ref{app:response-band}.

If only the absolute asymmetry $A$ is specified, its least required
allowance is $\min\{\epsilon_{\rm req}(A),\epsilon_{\rm req}(-A)\}$;
retaining the signed endpoint comparison supplies the stronger test.

\paragraph{Comparison using only response shape.}
For continuous positive responses on the same affine interval, define
$\delta=\osc\log(h/h_0)$. The signed means $\sigma,\sigma_0$
in \eqref{ctl:band-sharp} also obey the convenient, generally conservative
comparison
\begin{equation}
 |\operatorname{arctanh}\sigma-\operatorname{arctanh}\sigma_0|
 \le\frac\delta2,\qquad
 \inf_{c>0}\left\|\frac{h}{c h_0}-1\right\|_\infty
 =\tanh(\delta/2).
 \label{ctl:shape-distance}
\end{equation}
Each positive Bregman kernel ratio lies between the extrema of $h/h_0$;
their quotient proves the first assertion. The optimal normalization
is the arithmetic mean of those extrema. This shape comparison is
independent of positive scale, while physical energy remains calibrated.
Retaining the reference shape gives the stronger inverse
\eqref{ctl:inverse-shape}.

\subsection{Response bands conditioned on the energy increment}
\label{ctl:fixed-energy-sec}

Supplying the energy increment fixes the response integral discarded
by normalization, giving a linear moment optimization.
In the canonical calibration of section~\ref{phys:sec},
$H=\Delta E/\Delta\beta$ and $\alpha=-M/H$.
The following specialization of the bathtub principle
\cite[theorem 1.14]{liebloss2001} permits one-sided allowances.

\begin{proposition}[Response band with fixed integral]
\label{ctl:fixed-integral}
Use $s,h_0,H_0,M_0$ from proposition~\ref{ctl:sharp-band}. Fix
$0\le l<u$ and $H\in[lH_0,uH_0]$, and put
\[
 W=\frac{H-lH_0}{u-l}.
\]
Define $\tau_+,\tau_-\in[0,1]$ uniquely by
\begin{equation}
 \int_0^{\tau_+}h_0\,\dd t=W,
 \qquad \int_{\tau_-}^1h_0\,\dd t=W.
 \label{ctl:fixed-thresholds}
\end{equation}
For $lh_0\le h\le uh_0$ and $\int_0^1h=H$, the exact measurable
range of $M=\int_0^1s h$ is $[M_-,M_+]$, where
\begin{align}
 M_+&=lM_0+(u-l)\int_0^{\tau_+}s h_0\,\dd t,\nonumber\\
 M_-&=lM_0+(u-l)\int_{\tau_-}^1s h_0\,\dd t.
 \label{ctl:fixed-moments}
\end{align}
The unique extremizers, up to null sets, are respectively
$h_+=h_0\{l+(u-l)\mathbf1_{[0,\tau_+]}\}$ and
$h_-=h_0\{l+(u-l)\mathbf1_{[\tau_-,1]}\}$.

If $lH_0<H<uH_0$, continuous responses approach but do not attain the
endpoints, keeping $H$ \emph{exactly} fixed, and attain every interior
moment. The same holds for smooth responses when $h_0$ is smooth.
At $H=lH_0$ or $H=uH_0$, the sole response is respectively $lh_0$ or
$uh_0$ almost everywhere; hence $M=(H/H_0)M_0$ and the corresponding
continuous or smooth extremum is attained.
\end{proposition}

Responses are positive for $l>0$; $l=0$ serves only as a closed limiting
class below. For $H>0$, dividing \eqref{ctl:fixed-moments} by $H$ gives
the sharp signed-comparison interval, with thermodynamic calibration
\eqref{ctl:calibrated-moments}.

\begin{corollary}[Correction required by calibrated endpoint increments]
\label{ctl:calibrated-inverse}
For supplied $H>0$ and $M\in\mathbb R$, set
$d_H=H-H_0$, $d_M=M-M_0$, and
\begin{equation}
 \epsilon_{\rm cal}(H,M)
 =\max\left\{\frac{|d_H|}{H_0},\quad
 \max_{-1\le c\le1}\frac{|d_M-cd_H|}{J_0(c)}\right\}.
 \label{ctl:calibrated-epsilon}
\end{equation}
There exists a measurable response with these two moments and
$\|h/h_0-1\|_\infty\le\epsilon<1$ exactly when
$\epsilon_{\rm cal}\le\epsilon<1$.
Thus $\epsilon_{\rm cal}\ge1$ excludes every allowance below one.
For $\epsilon_{\rm cal}<1$, it is the attained measurable minimum and
the continuous infimum. Continuous attainment occurs exactly when
$M=(H/H_0)M_0$: the unique minimizer is $h=(H/H_0)h_0$ and
$\epsilon_{\rm cal}=|H/H_0-1|$. Otherwise the minimizer has one interior jump. Every larger allowance below one
admits a continuous response with \emph{both} moments exact.
The continuous statements also hold for smooth responses if $h_0$
is smooth.
\end{corollary}

Appendix~\ref{app:fixed-energy} proves these statements.
The calibrated inverse is at least the energy-only requirement. If
$|M|<H$, it is also at least the normalized inverse
\eqref{ctl:inverse-epsilon} at $m=M/H$, but their maximum need not be
sufficient. If $|M|\ge H$, no positive response has these moments.
These are constraints within the declared response class, not a claim
that all extremal responses have a gravitational realization.

\section{Finite entropy inference in a black-hole cavity}
\label{cav:sec}

A selected branch, reference response, same-temperature envelope and
calibrated energy increment give an entropy interval. The cavity
equations of state test inference from these reduced inputs; retaining
the exact family sharpens charge and entropy consistency tests.

\subsection{The Schwarzschild reference}
Fix the areal wall radius $R$ and its proper inverse temperature
$\beta_R$. For a Schwarzschild horizon with $z=r_+/R\in(0,1)$, the
flat-subtracted Brown-York energy and the equation of state are
\cite{york1986,brownyork1993,andrelemos2021}
\begin{equation}
 E_R=\frac RG(1-\sqrt{1-z}),\qquad
 S_R=\frac{\pi R^2z^2}{G},\qquad
 \beta_R=4\pi Rz\sqrt{1-z}.
 \label{cav:eos}
\end{equation}
All derivatives keep $R$ fixed, so $\dd S_R=\beta_R\dd E_R$. The
canonical branch potential and response are
\begin{equation}
 \psi_R=S_R-\beta_RE_R,\qquad
 \psi_R'=-E_R,\qquad
 g_R=-\frac{\dd E_R}{\dd\beta_R}
 =\frac1{4\pi G(3z-2)}.
 \label{cav:response}
\end{equation}
The positive branch $2/3<z<1$ excludes the inverse-temperature branch
singularity at $z=2/3$ and wall-horizon limit $z=1$. We use a compact
interior interval and require only local thermodynamic positivity.

Choose the reference endpoints $z_a=0.9$ and $z_b=0.99$, with
$\beta_i=\beta_R(z_i)$ and $\beta_b<\beta_a$. Then
\begin{equation}
 \frac{\beta_a}{R}\simeq3.5764517757,\qquad
 \frac{\beta_b}{R}\simeq1.2440706908,\qquad
 \alpha_R\simeq0.0524245038.
 \label{cav:inference-endpoints}
\end{equation}
These radii label $I=[\beta_b,\beta_a]$; other models are matched at
these temperatures, not necessarily these radii.
Writing $\ell=\sqrt{1-z}$ gives
\begin{equation}
 g_R\,\dd\beta_R=\frac RG\,\dd\ell,\qquad
 \beta_R(\ell)=4\pi R(\ell-\ell^3).
 \label{cav:uniform-lapse}
\end{equation}
The normalized response measure is therefore uniform in lapse.
Appendix~\ref{cav:appendix} evaluates its moments by cubic integration.

\subsection{An exact classical matter deformation}
\label{cav:rn-controlled}

For a gravitationally derived allowance, consider a Reissner-Nordstr\"om
black hole in the same Einstein-Maxwell cavity, at fixed charge within
each canonical family \cite[eqs.~(2.11) and (3.3)]{lundgren2008}.
Use the metric normalization $q^2/r^2$ and put
$z=r_+/R$, $\chi=(q/R)^2$. With
\begin{equation}
 f=(1-z)(1-\chi/z),\qquad A=1-\chi/z^2,
 \label{cav:rn-fA}
\end{equation}
the flat-subtracted Brown-York energy, entropy and proper inverse
wall temperature are
\begin{equation}
 E_q=\frac RG(1-\sqrt f),\qquad
 S_q=\frac{\pi R^2z^2}{G},\qquad
 \beta_q=\frac{4\pi Rz\sqrt f}{A}.
 \label{cav:rn-eos}
\end{equation}
The nonextremal domain is $\sqrt\chi<z<1$. Flat subtraction fixes the
energy convention without making neutral empty space a competing state
at nonzero fixed charge. At fixed $R,q$,
$\dd S_q=\beta_q\dd E_q$, so
$\psi_q=S_q-\beta_qE_q$ and $\psi_q'=-E_q$. Direct differentiation gives
\begin{equation}
 \mathcal D=zA^2-2f(1-3\chi/z^2),\qquad
 \frac{\dd\beta_q}{\dd z}
 =-\frac{2\pi R\mathcal D}{\sqrt f A^2},\qquad
 g_q=-\frac{\dd E_q}{\dd\beta_q}
 =\frac{A^3}{4\pi G\mathcal D}.
 \label{cav:rn-response}
\end{equation}
Equivalently $g_q=T_R^2 C_q$, where $T_R=1/\beta_q$ and
$C_q=(\partial E_q/\partial T_R)_{R,q}$ is the canonical heat capacity
\cite[eq.~(23)]{wang2020}. The derivatives hold charge and wall radius fixed.
We use $0\le|q|/R\le1/10$, uniquely matched on $z_q\ge9/10$.
Branch selection is an independent input: small and large
positive-response branches can coexist at the same charge and wall
temperature \cite[section VI]{lundgren2008}. The cutoffs $0.9$, $0.99$
and $|q|/R\le0.1$ provide a compact comparison separated from
extremality, the wall and loss of positive response; no optimality is
claimed. Separation bounds follow below and in
appendix~\ref{cav:rn-envelope-proof}.

\begin{proposition}[Uniform response control on the large-horizon branch]
\label{cav:rn-envelope}
Let $I$ be the proper inverse-temperature interval in
\eqref{cav:inference-endpoints}, labelled by the uncharged reference
radii $z_0\in[9/10,99/100]$. For every fixed
$0\le |q|/R\le1/10$ there is a unique solution of
$\beta_q(z_q)=\beta_R(z_0)$ with $9/10\le z_q<1$.
It has positive canonical response and obeys, uniformly on $I$,
\begin{equation}
 \begin{gathered}
 0\le z_q-z_0\le\frac{153}{80000},\qquad
 \lambda_{\rm RN}\le\frac{g_q(\beta_R)}{g_R(\beta_R)}\le1,\\
 \lambda_{\rm RN}
 =\left(\frac{80}{81}\right)^3\frac{2095200}{2080609}
 =\frac{39731200000}{40952626947}>\frac{97}{100}.
 \end{gathered}
 \label{cav:rn-uniform-envelope}
\end{equation}
\end{proposition}

Appendix~\ref{cav:rn-envelope-proof} proves this nonperturbative bound,
including the radius shift at fixed wall temperature:
$\|g_q/g_R-1\|_\infty\le1-\lambda_{\rm RN}=0.0298253626\ldots$.
The upper comparison is strict at nonzero charge, giving
$\Delta E_q<\Delta E_R$. Only the charge bound and selected branch,
not the fixed charge itself, need be known.

\begin{remark}[Branch selection is necessary]
\label{cav:small-branch}
Set $|q|/R=1/40$ and $1/40<z\le3/100$. Then
$0<A\le11/36$, $3\chi/z^2-1\ge13/12$, and $f'=-A<0$ gives
$f\ge4559/4800$. Thus $\mathcal D\ge59267/28800>0$:
this small branch also has positive response and decreasing $\beta_q$.
As $z\downarrow1/40$, $\beta_q\to\infty$, whereas
$[\beta_q(3/100)/\beta_b]^2=13677/14641<1$.
Every $\beta\in I$ therefore has a unique small-branch solution in this
radius interval. At its matched reference radius,
\begin{equation}
 \frac{g_{\rm small}(\beta)}{g_R(\beta)}
 =\frac{A^3(3z_0-2)}{\mathcal D}
 \le\frac{(11/36)^3(97/100)}{59267/28800}
 =\frac{1331}{98982}<0.014.
 \label{cav:small-branch-separation}
\end{equation}
Hence charge, wall temperature and local positivity alone do not imply
\eqref{cav:rn-uniform-envelope}. No global phase preference is assumed.
\end{remark}

Algebraically, write $g_q/g_R=c(1+e_{\rm RN}\xi)$, where
$|\xi|\le1$, $c=(1+\lambda_{\rm RN})/2$ and
\begin{equation}
 e_{\rm RN}=\frac{1-\lambda_{\rm RN}}{1+\lambda_{\rm RN}}
 =0.0151384360\ldots.
 \label{cav:rn-effective-band}
\end{equation}
The factor $c$ cancels from normalized comparisons; it does not
recalibrate $E_q$. Proposition~\ref{ctl:sharp-band} gives exact signed
limits $\alpha_-^{\rm RN}=-m_+(e_{\rm RN})$ and
$\alpha_+^{\rm RN}=-m_-(e_{\rm RN})$, with
\begin{equation}
 \alpha_-^{\rm RN}\le\alpha_q\le\alpha_+^{\rm RN},\qquad
 (\alpha_-^{\rm RN},\alpha_+^{\rm RN})
 \simeq(0.0447952129,\,0.0600478351).
 \label{cav:rn-alpha-inference}
\end{equation}
By \eqref{phys:secant-certificate}, the corresponding entropy secants obey
\begin{equation}
 s_-^{\rm RN}\le\frac{\Delta S_q}{R\Delta E_q}\le s_+^{\rm RN},\qquad
 (s_-^{\rm RN},s_+^{\rm RN})\simeq(2.4625009868,\,2.4802884507).
 \label{cav:rn-entropy-inference}
\end{equation}
These decimals evaluate exact root-defined bounds, sharp in the larger
one-sided response band with the regularity qualifications of
proposition~\ref{ctl:sharp-band}.

\subsection{The joint entropy-energy region}
\label{cav:inference}

Condition on the energy increment. For any positive canonical response
in the proved band, use $E=-\psi'$, $S=\psi+\beta_R E$ with the reference
wall variables and energy convention. Define
\begin{equation}
 \eta=\frac{\Delta E}{\Delta E_R},\qquad
 \nu=\frac{\Delta S}{R\Delta E_R},\qquad
 \lambda=\lambda_{\rm RN},\qquad
 w=\frac{\eta-\lambda}{1-\lambda}\in[0,1].
 \label{cav:region-coordinates}
\end{equation}
Put $u=\sqrt{0.1}$, $v=0.1$, $d=u-v$, and define the dimensionless
primitive and reference secant
\begin{equation}
 p(\ell)=4\pi\left(\frac{\ell^2}{2}-\frac{\ell^4}{4}\right),\qquad
 b(\ell)=p'(\ell)=\frac{\beta_R(\ell)}R,\qquad
 B=\frac{p(u)-p(v)}d.
 \label{cav:region-primitive}
\end{equation}
Then proposition~\ref{ctl:fixed-integral} gives the complete measurable
feasible region
\begin{align}
 \lambda\le\eta\le1,\qquad &\nu_-(\eta)\le\nu\le\nu_+(\eta),\nonumber\\
 \nu_-(\eta)&=\lambda B+(1-\lambda)\frac{p(v+wd)-p(v)}d,\nonumber\\
 \nu_+(\eta)&=\lambda B+(1-\lambda)\frac{p(u)-p(u-wd)}d.
 \label{cav:region-boundaries}
\end{align}
Since $g_R\,\dd\beta_R=(R/G)\,\dd\ell$, $\eta$ and $\nu$ are the uniform
lapse means of $r=g/g_R$ and $b(\ell)r$. At fixed $\eta$, the excess
$r-\lambda$ has mass $(1-\lambda)wd$; allocating it to the lowest or
highest lapses bounds its moment because $b$ increases.
The fixed-integral proposition proves sufficiency and exact-mass smooth
approximation: for $\lambda<\eta<1$, the boundaries are smooth-response
infimum and supremum, and every interior $\nu$ is attained smoothly.
At $\eta=\lambda$ or $1$, the response is proportional and $\nu=\eta B$.

The lower boundary is convex and the upper boundary concave, since
$\nu_-'(\eta)=b(v+wd)$ and $\nu_+'(\eta)=b(u-wd)$ with $b'>0$.
At fixed energy the sharp secant interval is
$[\nu_-(\eta)/\eta,\nu_+(\eta)/\eta]$.

\begin{figure}[!htbp]
 \centering
 \includegraphics[width=\linewidth]{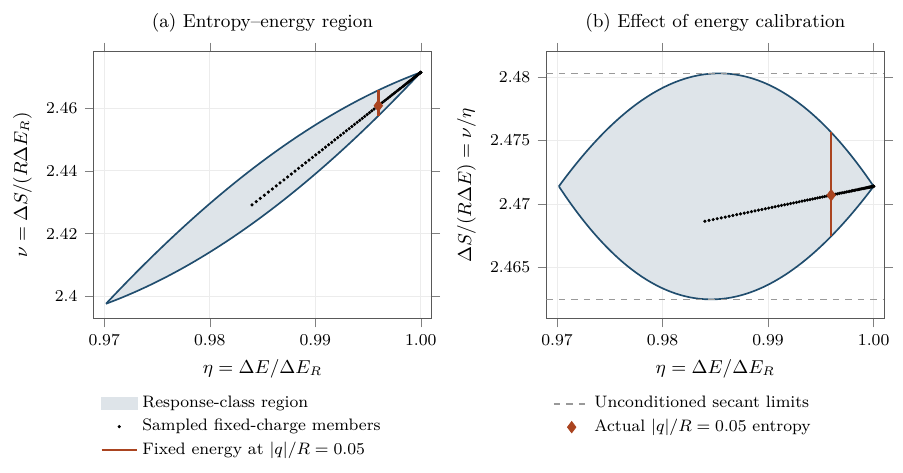}
 \caption{Entropy inference from the proved cavity response envelope.
 Left: the exact response-class region \eqref{cav:region-boundaries}.
 Right: its fixed-energy secant limits; dashed horizontal lines show the
 unconditioned limits \eqref{cav:rn-entropy-inference}. Dots sample 81
 fixed-charge large-branch solutions with $0\le|q|/R\le0.1$;
 proposition~\ref{cav:charge-ordering} proves their response ordering
 independently of these samples. The vertical segment marks the
 $|q|/R=0.05$ energy slice, and the diamond its actual entropy.}
 \label{fig:cavity-region}
\end{figure}

For a nonneutral calibration, let $\eta_*$ be the energy ratio obtained
from \eqref{cav:rn-eos} at the two matched large-branch radii with
$|q|/R=0.05$. Keeping only this energy ratio and the common
$|q|/R\le0.1$ envelope gives
\begin{equation}
 \eta_*\simeq0.9959774287,\qquad
 \left[\frac{\nu_-(\eta_*)}{\eta_*},\frac{\nu_+(\eta_*)}{\eta_*}\right]
 \simeq[2.4674664568,\,2.4756419242].
 \label{cav:nonneutral-calibration}
\end{equation}
The exact endpoint equations define $\eta_*$; its decimal is rounded.
The interval is about $54.0\%$ narrower than
\eqref{cav:rn-entropy-inference} under the same envelope.
The actual member has $\Delta S/(R\Delta E)\simeq2.4707038789$.
Figure~\ref{fig:cavity-region} separates the response-class boundaries
from sampled gravitational solutions.

Data outside \eqref{cav:region-boundaries} exclude the selected
large-branch family; data inside establish response-band compatibility,
not membership in that family.

\subsection{Charge inference within the exact gravitational family}
\label{cav:charge-identification}

Retaining the Einstein-Maxwell equations imposes correlations across
wall temperatures that a pointwise response band discards. Define
\begin{equation}
 \eta_{\rm RN}(\chi)=\frac{\Delta E_q}{\Delta E_R},\qquad
 \nu_{\rm RN}(\chi)=\frac{\Delta S_q}{R\Delta E_R},\qquad
 0\le\chi\le\chi_*=1/100,
 \label{cav:exact-family-coordinates}
\end{equation}
using the two matched large-branch endpoints.

\begin{proposition}[Charge ordering and endpoint identification]
\label{cav:charge-ordering}
On the family of proposition~\ref{cav:rn-envelope},
$(\partial_\chi g_q)_\beta<0$ at every fixed $\beta\in I$.
Consequently $\eta_{\rm RN}$ and $\nu_{\rm RN}$ are strictly decreasing.
An energy ratio in $[\eta_{\rm RN}(\chi_*),1]$ uniquely determines
$|q|/R$ and predicts $\nu$ within this family. If $\Xi$ is the inverse
of $\eta_{\rm RN}$, its exact entropy-energy curve is
$V(\eta)=\nu_{\rm RN}(\Xi(\eta))$, with
\begin{equation}
 \frac{\beta_b}{R}<V'(\eta)<\frac{\beta_a}{R}.
 \label{cav:exact-family-slope}
\end{equation}
\end{proposition}

Appendix~\ref{cav:charge-ordering-proof} proves the result by implicit
differentiation and uniform sign inequalities. Only charge magnitude is
identified because the equations depend on $q^2$. For an energy interval, set
$J=[\eta_L,\eta_U]\cap[\eta_{\rm RN}(\chi_*),1]$.
An empty $J$ excludes the selected family. Otherwise $J=[L,U]$ gives
\begin{equation}
 \Xi(U)\le\chi\le\Xi(L),\qquad
 V(L)\le\nu\le V(U).
 \label{cav:charge-inverse-interval}
\end{equation}
Both marginal intervals are exact and attained; simultaneous data must
lie on $\nu=V(\eta)$, rather than merely in the enclosing rectangle.
An additional entropy interval is compatible exactly when it intersects
$[V(L),V(U)]$. The slope bound controls energy-uncertainty propagation:
$V(U)-V(L)\le(\beta_a/R)(U-L)$.

For an entropy-discriminating test, take
$\eta_*=\eta_{\rm RN}(1/400)$ from \eqref{cav:nonneutral-calibration}.
The smooth proportional response $g=\eta_*g_R$ lies in the proved band
and has an energy ratio admitted by the exact family. Nevertheless,
\begin{equation}
 \nu_{\rm prop}=\eta_*B\simeq2.4614568184
 \;>\;V(\eta_*)=\nu_{\rm RN}(1/400)\simeq2.4607652965.
 \label{cav:entropy-exclusion}
\end{equation}
The dimensionless gap is $\nu_{\rm prop}-V(\eta_*)\simeq6.9152\times10^{-4}$.
The exact-rational sign certificate in
appendix~\ref{cav:entropy-exclusion-proof} excludes this pair independently
of rounding: its energy is admissible, but its entropy violates the
Einstein-Maxwell correlation. These theoretical tests require calibrated
quasilocal energies; no observational protocol is assumed.
The inverse in $\chi$ is smooth at neutrality, whereas charge magnitude
has square-root sensitivity there because $\eta_{\rm RN}'(0)<0$.

\subsection{Comparing geometric certificates on the same envelope}

The reference budgets on the chosen interval are
\begin{equation}
 \begin{gathered}
 \omega_R=\Jvar_R=\log(97/70),\qquad K_R=\min\{\kappa_R,\kappa_E\},\\
 \kappa_R\simeq0.7186953204,\qquad \kappa_E\simeq0.5860904863.
 \end{gathered}
 \label{cav:compact-budgets}
\end{equation}
Appendix~\ref{cav:compact-slopes} gives the two slope maxima.
With $e=e_{\rm RN}$, the joint transfer gives
$T_e[\mathcal F(\omega_R,K_R)]\simeq0.088006$, compared with the
direct range bound
$\tanh[(\omega_R+\log(1/\lambda_{\rm RN}))/4]\simeq0.088889$.
The improvement is about $0.99\%$; equation~\eqref{ctl:joint-advantage}
specifies its response-error budget. Full reference-shape information
gives the tighter band bound $0.060048$ in
\eqref{cav:rn-alpha-inference}. These certificates use progressively
different information; the exact-family inverse instead retains the
field-equation correlations.

Each directed divergence uses one fixed-charge potential, implying no
relative entropy between charge sectors. Positive response establishes
local thermodynamic positivity, not global phase preference, control of
other matter phases or radiation, or microscopic saddle accuracy.
These reduced-information certificates complement direct equation-of-state
solutions and apply to further responses with independently justified
same-temperature envelopes.

\section{Conclusions}
\label{sec:discussion}
\label{sec:conclusion}

The sharp joint range-slope theorem bounds directed comparisons
from a positive scalar response. Its invariant form combines both
Legendre slope budgets. It applies to exact quantum Bregman
contrasts and, with first-law calibration, to finite thermodynamic
Clausius deficits.

For nested thermal CFT$_2$ intervals, a positive Laplace
representation gives strict log-convexity of the loss response.
The finite-step asymmetry then has an exact monotone inverse and
predicts contrasts across accessible fractions, independently of
central-charge amplitude. A convergent dual-budget enclosure makes
the full Legendre certificate computable; the cubic is the
small-step limit. With calibrated interval sizes and affine
increments, the semiclassical planar-BTZ dictionary converts the
inverse into boundary-normalized thermal and exterior depth scales.
Estimating the contrasts still requires a measurement protocol,
and sensitivity decreases at high temperature.

In the non-AdS cavity, the selected large-horizon branch obeys
$\lambda_{\rm RN}g_R\le g_q\le g_R$ uniformly; a coexisting small
positive-response branch violates this envelope. Energy calibration
narrows the response-class entropy interval. In the exact
Einstein-Maxwell family, charge ordering lets energy determine
charge magnitude and predict entropy, with attained inverse
intervals. The proportional-response example passes the family's
energy-range test but fails its entropy prediction, separating
response-band compatibility from exact-family membership.

Thermodynamics alone does not specify quantum access.
For nonscalar finite-dimensional $h$, the Hamiltonians $h\otimes I$
and $I\otimes h$ on identical factors have the same partition
function and input thermal contrasts. Tracing the second factor
preserves those contrasts for the first model and erases them for
the second. The cavity potential therefore supplies neither a
subsystem algebra nor an accessibility map. A scalar cubic likewise
supplies no intrinsic non-Abelian holonomy; the companion note's
compact connection and three-level channels need additional
carriers and physical dynamics to describe a black hole.

Sharpness concerns the declared response classes, not the smaller
CFT or Einstein-Maxwell families. Controlled deformations can be
tested against the certificates if they retain the contrast
structure and admit a uniform response envelope. Otherwise the
contrast and its derivatives require separate control.
These equilibrium comparisons specify neither dissipative
evolution, global saddle preference nor behind-horizon reconstruction.

\clearpage
\section*{Appendices}
\appendix
\section{Proof of the joint response comparison}
\label{app:joint-proof}

\begin{proof}[Proof of theorem~\ref{bnd:joint-theorem}]
The case $\omega=0$ is constant. For $\omega>0$, put
\begin{equation}
 R=e^\omega,\quad D=R-1,\quad \ell=\omega/\kappa,\quad
 c=\frac{D-\omega}{\kappa D}.
 \label{bnd:joint-parameters}
\end{equation}
Here $0<c<\ell$ is the mean of $z$ under $e^{-\kappa z}$ on
$[0,\ell]$. For the profile $h_s$ in \eqref{bnd:joint-extremizer},
with $0\le s\le1-\ell$, its ramp mean is $\mu_s=s+c$. Set
\begin{equation}
 F(s)=\int_0^1(\mu_s-t)h_s(t)\,\dd t.
 \label{bnd:joint-placement}
\end{equation}
The ramp contributes zero. If $\ell<1$, the late plateau gives
$F(0)<0$, the early plateau gives $F(1-\ell)>0$, and
\begin{equation}
 F'(s)=\int_0^1h_s\,\dd t
   +\kappa\int_s^{s+\ell}(\mu_s-t)h_s\,\dd t
 =\int_0^1h_s\,\dd t>0.
 \label{bnd:joint-placement-unique}
\end{equation}
Thus a unique $s_*\in(0,1-\ell)$ matches the full and ramp means.
For $\ell=1$, the only placement is $s_*=0$ and
$F(0)=0$. Write $h_*=h_{s_*}$ and $\mu=s_*+c$.

Normalize any admissible competitor to $1\le h\le R$. The early
plateau has $\mu-t>0$ and $h\le h_*$; the late plateau has
$\mu-t<0$ and $h\ge h_*$. Hence both give
$(\mu-t)h\le(\mu-t)h_*$. On the ramp $I=[s_*,s_*+\ell]$,
$q=h/h_*$ is nondecreasing because
\begin{equation}
 (\log q)'=(\log h)'+\kappa\ge0\quad\hbox{almost everywhere}.
 \label{bnd:joint-ramp-order}
\end{equation}
For any probability density $w$,
\begin{equation}
 \Cov_w(t,q)=\frac12\int\!\!\int
 (t-r)\{q(t)-q(r)\}w(t)w(r)\,\dd t\,\dd r.
 \label{bnd:slope-covariance}
\end{equation}
Taking $w\propto h_*\mathbf1_I$, whose mean is $\mu$, gives
\begin{equation}
 \int_I(\mu-t)h\,\dd t\le0
 =\int_I(\mu-t)h_*\,\dd t.
 \label{bnd:joint-certificate}
\end{equation}
Including the plateaus gives $\int(\mu-t)h\le0$, so the mean of $h$
is at least $\mu$ and, by reflection, at most $1-\mu$.
Thus $\mathcal A[h]\le1-2\mu$, attained by $h_*$ and its reflection,
without assuming monotonicity of $h$.

Direct integration gives, for
$H_j(s)=\int_0^1t^j h_s(t)\,\dd t$,
\begin{align}
 H_0(s)&=A+Ds,\qquad A=1+(D-\omega)/\kappa,\nonumber\\
 H_1(s)&=Ds^2/2+(D-\omega)s/\kappa
       +1/2+(D-\omega-\omega^2/2)/\kappa^2.
 \label{bnd:joint-moments}
\end{align}
Solving $(s+c)H_0(s)=H_1(s)$ at its unique admissible root yields
\begin{equation}
 W=R+\frac{D^2-R\omega^2}{\kappa^2},\qquad
 s_* =\frac{\sqrt W-A}{D},\qquad
 \mu=\frac{\sqrt W-1}{D}.
 \label{bnd:joint-root}
\end{equation}
Then $1-2\mu=(R+1-2\sqrt W)/D$ is \eqref{bnd:joint-formula}.

If $0<\omega<\kappa$, equality forces the two plateau values.
Covariance equality forces $q$ constant: a nonconstant continuous
nondecreasing $q$ makes the integrand positive on a set of positive
product measure. Continuity at the joins fixes
$q=1$. The optimizer and its reflection are therefore unique up to scale;
their derivative jumps exclude $C^1$ attainment.
For $\omega=\kappa$, the ramp is the whole interval. The same strict
covariance argument gives $h=ch_*$, proving the exponential equality
case directly.

For $0<\omega<\kappa$, extend $\log h_*$ by its constant tails and
convolve with a nonnegative smooth unit-mass mollifier. Sufficiently
small support retains both plateau values, monotonicity and the slope
bound $\kappa$. Exponentiation gives smooth responses with range
exactly $\omega$, converging uniformly to $h_*$. Their positive denominators
ensure convergence of $\mathcal A$.
Integrating twice gives smooth positive-Hessian potentials. The cases
$\omega=\kappa$ and $\omega=0$ are already smooth.

The boundary values follow from the formula. Since
$2\sinh(\omega/2)>\omega$ for $\omega>0$, the finite-slope value is
strictly below $\tanh(\omega/4)$. When $\omega<\kappa$, equality in
the separate slope bound would require an exponential of range
$\kappa$, excluded by the smaller range budget. Monotonicity follows
from nested admissible classes.

The minimum affine budget also retains sharpness.
For $h_*$ with minimum one, $\kappa_{\rm p}=\kappa$ and
\[
 \kappa_{\rm d}=\left(\int_0^1h_*\right)
       \mathop{\rm ess\,sup}\frac{|h_*'|}{h_*^2}
 =\kappa\int_0^1h_*>\kappa.
\]
For $\omega<\kappa$, sufficiently narrow logarithmic mollifications
retain a central ramp of slope $\kappa$. Their suprema of $|h'|/h^2$ tend to
$\kappa$: they are at most $\kappa$, and the retained ramp reaches
response values tending to one. Their integrals converge too, so
$\kappa_{\rm d}>\kappa_{\rm p}=\kappa$ eventually and $K=\kappa$.
For $\omega=\kappa$, the exponential already has these budgets.
This proves the claimed smooth supremum for the smaller affine slope.
\end{proof}

\section{Sharp response bands and inverse corrections}
\label{app:response-band}

\subsection{Extremal signed comparisons}
The notation is that of proposition~\ref{ctl:sharp-band}.
\begin{proof}
For fixed $m$, maximizing or minimizing $\int(s-m)h$ pointwise gives
$M_0-mH_0\pm\epsilon J_0(m)$. These functions are continuous and
strictly decreasing: their decrease between $m$ and $m+d$, $d>0$, is
at least $(1-\epsilon)H_0d$. They are positive at $-1$ and negative at
$1$. At their roots the extremizing responses are
\begin{equation}
 h_\pm(t)=h_0(t)\{1\pm\epsilon\,\operatorname{sgn}[s(t)-m_\pm]\}.
 \label{ctl:band-extremizers}
\end{equation}
Dividing by $\int h>0$ proves the signed bounds and equality.
For $\epsilon>0$, equality forces the displayed response almost
everywhere except at its interior threshold; its nonzero jump precludes
continuity. Replace the sign by $\tanh[n(s-m_\pm)]$ and use dominated
convergence to obtain continuous, or smooth, admissible approximants.
The final comparison follows from theorem~\ref{ctl:transfer}, applied
to these approximants.
\end{proof}

\subsection{Minimum correction for a normalized endpoint comparison}
The notation is that of corollary~\ref{ctl:inverse-band}.
\begin{proof}
The constraint $\int(s-m)h=0$ and pointwise error bound give
$|mH_0-M_0|\le\epsilon J_0(m)$. Both signs of $s-m$ occur on sets of
positive measure, so this quotient is strictly below one. Equality is
realized by
\[
 \frac h{h_0}=1+\frac{mH_0-M_0}{J_0(m)}\operatorname{sgn}(s-m).
\]
A nonzero coefficient gives an interior jump, uniquely forced almost
everywhere by equality. Replacing the sign by
$\tanh[n(s-m)]$ and replacing the coefficient by
\[
 \frac{mH_0-M_0}{\int_0^1(s-m)\tanh[n(s-m)]h_0\,\dd t}
\]
keeps the mean exactly $m$ and approaches the minimum allowance from
above. A ratio with positive lower and upper bounds $q_-,q_+$ can be scaled into a symmetric
relative band of half-width
$(q_+-q_-)/(q_++q_-)=\tanh[\tfrac12\log(q_+/q_-)]$.
Conversely, the displayed extremizer has exactly that ratio of upper
and lower values. Normalized means are unchanged by this scaling.
\end{proof}

\subsection{Evaluation from the reference potential}
Set $\Delta=b-a$ and $h_0(t)=\psi_0''(a+t\Delta)$.
No response quadrature is needed when $\psi_0$ and $\psi_0'$ are known. Let
$D_0=B_{\psi_0}(b,a)-B_{\psi_0}(a,b)$,
$\Sigma_0=B_{\psi_0}(b,a)+B_{\psi_0}(a,b)$ and
$c_m=a+(1-m)\Delta/2$. Taylor's integral remainder gives
\begin{equation}
 \Delta^2 M_0=D_0,\quad \Delta^2H_0=\Sigma_0,\quad
 \Delta^2J_0(m)=4B_{\psi_0}(c_m,a)-D_0+m\Sigma_0.
 \label{ctl:band-endpoint-form}
\end{equation}

\subsection{Fixed response integral and calibrated inverse}
\label{app:fixed-energy}

\begin{proof}[Proof of proposition~\ref{ctl:fixed-integral}]
Write $h=h_0\{l+(u-l)f\}$, so $0\le f\le1$ and
$\int f h_0=W$. For $f_+=\mathbf1_{[0,\tau_+]}$,
\[
 (s-s(\tau_+))(f-f_+)\le0.
\]
Equal masses give $\int s f h_0\le\int s f_+h_0$ after integration
against $h_0$. Strict monotonicity of $s$ forces
$f=f_+$ almost everywhere in equality. Reversing the inequality with
$f_-=\mathbf1_{[\tau_-,1]}$ gives the minimum. Convex combinations
of the two extremizers fill the entire moment interval. If $W=0$ or
$W=H_0$, the integral and $0\le f\le1$ force $f=0$ or $f=1$.

For $0<W<H_0$, replace the upper indicator by
\[
 f_{n,+}(t)=\frac1{1+\exp[-n(s(t)-c_{n,+})]},
 \qquad \int f_{n,+}h_0=W.
\]
A unique real $c_{n,+}$ exists because the integral decreases
continuously and strictly from $H_0$ to zero with the threshold.
Its limit is $s(\tau_+)$: evaluating the integral at either side of
that value and using dominated convergence brackets the threshold.
Thus $f_{n,+}$ converges almost everywhere to $f_+$ and preserves mass
exactly. For the lower extremizer use
$[1+\exp\{n(s-c_{n,-})\}]^{-1}$, with its threshold again chosen to
fix the mass. The responses are continuous, smooth when $h_0$ is smooth,
and have first moments converging to $M_+$ and $M_-$.
The interior jumps and equality classification exclude continuous
endpoint attainment. Given any $M_-<M<M_+$, sufficiently advanced
upper and lower approximants have moments on opposite sides of $M$;
their convex combination with coefficient
$(M-M_{n,-})/(M_{n,+}-M_{n,-})$ has exactly the prescribed $H$ and $M$.
\end{proof}

\begin{proof}[Proof of corollary~\ref{ctl:calibrated-inverse}]
Necessity follows by integrating $h-h_0$ and $(s-c)(h-h_0)$ against
the relative envelope:
\[
 |d_H|\le\epsilon H_0,\qquad
 |d_M-cd_H|\le\epsilon J_0(c).
\]
For sufficiency, suppose these inequalities hold and set
$l=1-\epsilon$, $u=1+\epsilon$. If $\epsilon=0$, they force $(H,M)=(H_0,M_0)$.
Otherwise the first inequality puts $H$ in the allowed mass interval.
The fixed-mass upper and lower bounds have the equivalent forms
\begin{align}
 M_+(H,\epsilon)
 &=\min_{-1\le c\le1}\{M_0+cd_H+\epsilon J_0(c)\},\nonumber\\
 M_-(H,\epsilon)
 &=\max_{-1\le c\le1}\{M_0+cd_H-\epsilon J_0(c)\}.
 \label{ctl:calibrated-support}
\end{align}
Indeed the pointwise bound on $\int(s-c)(h-h_0)$ gives one direction;
choosing $c=s(\tau_+)$ or $s(\tau_-)$ from the fixed-mass extremizer
makes it equality. The assumed inequalities put $M$ between these bounds;
proposition~\ref{ctl:fixed-integral} then supplies a measurable response
with both moments. Since $J_0>0$ on the compact
interval, the maximum in \eqref{ctl:calibrated-epsilon} exists, proving
the attained measurable minimum whenever it is below one.

Put $\epsilon_H=|H/H_0-1|$. At $\epsilon=\epsilon_H$ the mass constraint
forces $h=(H/H_0)h_0$. If $M\ne(H/H_0)M_0$, the minimum is strictly
larger than $\epsilon_H$. At this minimum $M$ must be an endpoint of
the fixed-mass moment interval; otherwise continuity would allow a
smaller budget. Its extremizer then has an interior jump. Conversely,
proportional moments attain $\epsilon_H$ uniquely by the constant ratio.

The endpoint intervals widen strictly with budget.
For $\epsilon>\epsilon_H$, differentiating
$W=H_0/2+d_H/(2\epsilon)$ in \eqref{ctl:fixed-moments} gives
\begin{equation}
 \partial_\epsilon M_+(H,\epsilon)=J_0(s(\tau_+))>0,
 \qquad
 \partial_\epsilon M_-(H,\epsilon)=-J_0(s(\tau_-))<0.
 \label{ctl:calibrated-monotonicity}
\end{equation}
Thus at every larger budget below one, both the mass and target moment
are interior. The exactly mass-preserving approximants and their convex
combination constructed above match both moments, proving the stated
continuous and smooth infimum and exact-feasibility assertions.
\end{proof}

\section{Cavity response calculations}
\label{cav:appendix}

At fixed $R$, differentiating \eqref{cav:eos} gives
\begin{equation}
 \frac{\dd E_R}{\dd z}=\frac R{2G\sqrt{1-z}},\qquad
 \frac{\dd\beta_R}{\dd z}
 =-\frac{2\pi R(3z-2)}{\sqrt{1-z}}.
 \label{cav:first-derivatives}
\end{equation}
These give \eqref{cav:response} and \eqref{cav:uniform-lapse}. For $u=\sqrt{1-z_a}$ and
$v=\sqrt{1-z_b}$, the reference increments are
\begin{equation}
 \Delta E_R=\frac RG(u-v),\qquad
 \Delta S_R=\frac{\pi R^2}{G}(u-v)(u+v)(2-u^2-v^2).
 \label{cav:endpoint-differences}
\end{equation}
Substitution into \eqref{phys:signed-comparison} yields
$\alpha_R=(u^2-v^2)/[2(1-u^2-uv-v^2)]$ and the value in
\eqref{cav:inference-endpoints}.

Let $\Delta\beta=\beta_a-\beta_b$ and
$h_0(t)=g_R(\beta_b+t\Delta\beta)$. With
\begin{equation}
 P(\ell)=4\pi R\left(\frac{\ell^2}{2}-\frac{\ell^4}{4}\right),\qquad
 \langle\beta_R\rangle_0=\frac{P(u)-P(v)}{u-v},
 \label{cav:reference-primitive}
\end{equation}
the normalized moments are
\begin{equation}
 \begin{gathered}
 H_0=\frac{R(u-v)}{G\Delta\beta},\qquad
 \frac{M_0}{H_0}=-\alpha_R,\\
 \frac{J_0(m)}{H_0}
 =\frac1{u-v}\int_v^u
 \left|1-\frac{2[\beta_R(\ell)-\beta_b]}{\Delta\beta}-m\right|\dd\ell.
 \end{gathered}
 \label{cav:reference-moments}
\end{equation}
The last integral has one threshold, since $\beta_R(\ell)$ increases
on $[v,u]$. It solves
$\beta_R(\ell_m)=(\beta_a+\beta_b-m\Delta\beta)/2$.
Splitting there and using $P$ evaluates the integral exactly; the
monotone equations in proposition~\ref{ctl:sharp-band} determine the roots.

\subsection{The compact range-slope comparison}
\label{cav:compact-slopes}
Since $g_R$ decreases with radius, its logarithmic range and variation
coincide:
\begin{equation}
 \Jvar_R=\log\frac{3z_b-2}{3z_a-2}.
 \label{cav:variation}
\end{equation}
Differentiating \eqref{cav:response} along the regular $\beta_R$ branch
gives
\begin{equation}
 \frac{|g_R'|}{g_R}
 =\frac{3\sqrt{1-z}}{2\pi R(3z-2)^2},\qquad
 \frac{|g_R'|}{g_R^2}
 =\frac{6G\sqrt{1-z}}{R(3z-2)}.
 \label{cav:slope-functions}
\end{equation}
Both decrease strictly with $z$, as their numerator decreases and
positive denominators increase; their maxima occur at $z_a$. The
primal and dual affine lengths are respectively
$\Delta\beta$ and $\Delta E_R=R(u-v)/G$, so
\begin{equation}
 \kappa_R=\Delta\beta\frac{3u}{2\pi R(3z_a-2)^2},\qquad
 \kappa_E=\frac{6u(u-v)}{3z_a-2}.
 \label{cav:slope-budgets}
\end{equation}
At the reference endpoints these give \eqref{cav:compact-budgets}.
The joint certificate uses $K_R=\min\{\kappa_R,\kappa_E\}$.

\subsection{Proof of the uniform charged-cavity response envelope}
\label{cav:rn-envelope-proof}

Set $a=9/10$, $\chi_*=1/100$, $d=3z-2$, and let
$z\in[a,1)$, $0\le\chi\le\chi_*$. The response denominator expands as
\begin{equation}
 \mathcal D=d-2\chi+\frac{6\chi(1-z)}{z^2}
             +\frac{\chi^2(6z-5)}{z^3}.
 \label{cav:rn-denominator-expansion}
\end{equation}
The last two terms are nonnegative on this domain, so
$\mathcal D\ge d-2\chi\ge17/25>0$.
Also $(1-z)/z^2\le(1-a)/a^2$ and
$0<(6z-5)/z^3\le1$. Therefore
\begin{equation}
 \mathcal D-d
 \le-c\chi,\qquad
 c=2-\frac{6(1-a)}{a^2}-\chi_* =\frac{3373}{2700}.
 \label{cav:rn-denominator-upper}
\end{equation}
Since $c>0$, these inequalities prove positive response and strict
decrease of $\beta_q(z)$ on $[a,1)$.

To compare the temperature functions, put
\begin{equation}
 F_\chi(z):=\frac{\beta_q(z)}{\beta_R(z)}
 =\frac{\sqrt{1-\chi/z}}{1-\chi/z^2}.
 \label{cav:rn-temperature-ratio}
\end{equation}
The difference between the numerator squared and the denominator
squared is $\chi\{z^2(2-z)-\chi\}/z^4\ge0$, so $F_\chi\ge1$.
Moreover, concavity of the square root gives
\begin{equation}
 F_\chi(z)\le
 \frac{\sqrt{1-\chi}}{1-\chi/a^2}
 \le\frac{1-\chi/2}{1-\chi/a^2}.
 \label{cav:rn-temperature-upper}
\end{equation}
Because $\beta_q(a)\ge\beta_R(a)$ and $\beta_q(z)\to0$ as $z\to1$,
strict monotonicity gives a unique match $z_q\in[a,1)$ for every
$z_0\in[a,99/100]$. At $z=z_0$ the charged inverse temperature is
at least its reference value, so $z_q\ge z_0$.

The uncharged derivative magnitude
$|\beta_R'(z)|=2\pi R(3z-2)/\sqrt{1-z}$ increases on $[a,1)$.
The mean-value theorem and the matching equation yield
\begin{align}
 |\beta_R'(a)|(z_q-z_0)
 &\le\beta_R(z_0)-\beta_R(z_q)\nonumber\\
 &=\beta_R(z_q)\{F_\chi(z_q)-1\}\nonumber\\
 &\le\beta_R(a)
       \left\{\frac{1-\chi/2}{1-\chi/a^2}-1\right\}.
\end{align}
As $\beta_R(a)/|\beta_R'(a)|=2a(1-a)/(3a-2)$,
\begin{equation}
 0\le z_q-z_0\le
 \Delta(\chi):=
 \frac{a(1-a)\chi(2-a^2)}{(3a-2)(a^2-\chi)}
 \le\Delta(\chi_*)=\frac{153}{80000}.
 \label{cav:rn-shift-proof}
\end{equation}
All matched radii are at most $99/100+153/80000=0.9919125<1$,
uniformly away from the wall, extremality and loss of positive response.

For the upper response comparison, an exact rearrangement gives
\begin{equation}
 \mathcal D-dA^3
 =\chi\frac{3-2z}{z}
  +\chi^2\frac{6z^2-14z+6}{z^4}
  +\frac{d\chi^3}{z^6}.
 \label{cav:rn-response-order-identity}
\end{equation}
On $[a,1)$, $(3-2z)/z\ge1$ and
$6z^2-14z+6\ge-2$. Hence
\begin{equation}
 \mathcal D-dA^3
 \ge\chi\left(1-\frac{2\chi_*}{a^4}\right)\ge0.
\end{equation}
Consequently $g_q(z)\le[4\pi G(3z-2)]^{-1}=g_R(z)$ at the same radius.
Since $g_R$ decreases with radius and $z_q\ge z_0$, the matched ratio
is at most one.

For the lower comparison, retain the negative charge term in
\eqref{cav:rn-denominator-upper}. Since $\Delta(\chi)/\chi$ increases
with $\chi$, equation~\eqref{cav:rn-shift-proof} gives
$z_q-z_0\le(153/800)\chi$, including $\chi=0$ by continuity. With
$d_0=3z_0-2\in[7/10,97/100]$ and $k=c-459/800=14591/21600$,
\begin{equation}
 0<\mathcal D(z_q,\chi)\le d_0-k\chi,
 \qquad d_0-k\chi\ge\frac7{10}-\frac{k}{100}>0.
 \label{cav:rn-sharper-denominator}
\end{equation}
Because $A(z_q)\ge1-\chi/a^2>0$, this implies
\begin{align}
 \frac{g_q(\beta_R(z_0))}{g_R(\beta_R(z_0))}
 &=\frac{A(z_q)^3d_0}{\mathcal D(z_q,\chi)}\nonumber\\
 &\ge\left(1-\frac\chi{a^2}\right)^3\frac{d_0}{d_0-k\chi}
 \ge\left(1-\frac\chi{a^2}\right)^3
       \frac{97/100}{97/100-k\chi}.
 \label{cav:rn-response-lower-proof}
\end{align}
The second inequality uses the nonpositive derivative of
$d_0/(d_0-k\chi)$ with respect to $d_0$.
The final expression decreases with $\chi$: its logarithmic derivative
is negative precisely when $k(a^2+2\chi)<3(97/100)$, which holds since
$k(a^2+2\chi_*)=1211053/2160000<291/100$.
Its value at $\chi_*=1/100$ is
\begin{equation}
 \left(\frac{80}{81}\right)^3\frac{2095200}{2080609}
 =\frac{39731200000}{40952626947}=\lambda_{\rm RN}.
\end{equation}
This proves proposition~\ref{cav:rn-envelope} uniformly in proper wall
temperature and the unknown-but-fixed charge. The response-band extrema
and inverse tests follow from the one-dimensional comparisons.

\subsection{Proof of charge ordering and endpoint identification}
\label{cav:charge-ordering-proof}

Set $d=3z-2$ and $C=1-\chi d/z^3$.
Differentiating the equation of state at fixed proper wall temperature gives
\begin{equation}
 \left(\frac{\partial E_q}{\partial\chi}\right)_{\beta}
 =\frac RG\frac{\sqrt f\,C}{\mathcal D}.
 \label{cav:fixed-temperature-charge-derivative}
\end{equation}
Indeed, implicit differentiation first gives
$z_\chi=(1-z)(2z-z^2-\chi)/(z^2\mathcal D)$, and substitution into
$E_\chi+E_z z_\chi$ yields the displayed expression.
On the established domain $z\ge a=9/10$, $\chi\le1/100$,
\begin{gather*}
 f_z=-A<0,\qquad C\ge1-\chi\ge99/100,\qquad
 C_z=-6\chi(1-z)/z^4\le0,\\
 \mathcal D_z=3+\frac{6\chi(z-2)}{z^3}
 +\frac{\chi^2(15-12z)}{z^4}
 \ge3-\frac{6\chi_*(2-a)}{a^3}=\frac{707}{243}>0.
\end{gather*}
Here $(3z-2)/z^3\le1$ follows from its nonnegative derivative
$6(1-z)/z^4$ and its limit at the wall. Thus the right-hand side of
\eqref{cav:fixed-temperature-charge-derivative} decreases strictly with
$z$. Since $\beta_z<0$, smooth implicit differentiation gives
\begin{equation}
 \left(\frac{\partial g_q}{\partial\chi}\right)_\beta
 =-\partial_\beta
  \left(\frac{\partial E_q}{\partial\chi}\right)_\beta<0.
 \label{cav:charge-response-monotonicity}
\end{equation}
The identities $\Delta E_q=\int_I g_q\,\dd\beta$ and
$\Delta S_q=\int_I\beta g_q\,\dd\beta$, with $\beta>0$, prove both
strict monotonicity claims. Compactness and the positive denominator
justify differentiation under the integral.

Finally, with $w_\chi(\beta)=-(\partial_\chi g_q)_\beta>0$,
\begin{equation}
 V'(\eta)=\frac{\nu_{\rm RN}'(\chi)}{\eta_{\rm RN}'(\chi)}
 =\frac{\int_I(\beta/R)w_\chi(\beta)\,\dd\beta}
        {\int_I w_\chi(\beta)\,\dd\beta},\qquad
 \chi=\Xi(\eta).
\end{equation}
This strictly positive weighted mean lies between the two endpoint
inverse temperatures divided by $R$, proving
\eqref{cav:exact-family-slope}. Continuous monotone inversion gives
\eqref{cav:charge-inverse-interval} and its attainment. At the parameter
endpoints derivatives have their one-sided interpretation.

\subsection{An energy-admissible, entropy-excluded pair}
\label{cav:entropy-exclusion-proof}

Set $a=9/10$, $b=99/100$, $\chi=1/400$ and $d=\sqrt{1/10}-1/10$.
For a reference endpoint $z_0\in\{a,b\}$, the matched radius is the unique
large-branch root of
\[
 P(z;z_0)=z^5(1-z)(z-\chi)-z_0^2(1-z_0)(z^2-\chi)^2=0.
\]
This is the squared temperature equation; all factors squared are
positive on the selected branch. Rational substitution gives
\[
 \begin{split}
 0.900435389&<z_{q,a}<0.900435390,\\
 0.990026290&<z_{q,b}<0.990026291.
 \end{split}
\]
The lower endpoint has $P>0$ and the upper endpoint $P<0$ in each
bracket. For $f=(1-z)(1-\chi/z)$, direct rational squaring gives
\[
 \sqrt{f(z_{q,a})}>0.3151002,\qquad
 \sqrt{f(z_{q,b})}<0.0997423,\qquad d<0.2162278.
\]
Hence $\eta_*=[\sqrt{f(z_{q,a})}-\sqrt{f(z_{q,b})}]/d
>(0.3151002-0.0997423)/0.2162278>0.99597$. Consequently,
\[
 \eta_*(b^2-a^2)-(z_{q,b}^2-z_{q,a}^2)
 >0.99597(b^2-a^2)-(0.990026291^2-0.900435389^2)
 >\frac1{25000}.
\]
Multiplying by $\pi/d>0$ gives
$\nu_{\rm prop}-\nu_{\rm RN}(1/400)>0$, proving the strict entropy
separation in \eqref{cav:entropy-exclusion}.

\setlength{\bibsep}{6pt}
\renewcommand{\bibfont}{\interlinepenalty=10000}
\bibliographystyle{iopart-num-cqg}
\bibliography{references}
\end{document}